\documentstyle[preprint,aps,epsfig,psfig]{revtex}
      \newcommand{\beq}{\begin{eqnarray}}
      \newcommand{\eeq}{\end{eqnarray}}

       \def\sl{\slash{\hskip -2.5mm}}
      \def\sla{\slash{\hskip -2mm}}

\begin{document}

      \draft

      \title{QCD factorization for
      $B \to PP$ \footnote{Supported in part by
     National Natural Science Foundation of China and State Commission of
      Science and Technology of China}}
     \vspace{2cm}

      \author{ Dongsheng Du${}^{1,2}$, Deshan Yang${}^{2}$ and Guohuai
      Zhu${}^{2}$ \footnote{Email: duds@mail.ihep.ac.cn,
      yangds@mail.ihep.ac.cn, zhugh@mail.ihep.ac.cn} } \address{${}^1$ CCAST
      (World Laboratory), P.O.Box 8730, Beijing 100080, China\\ ${}^2$ Institute
      of High Energy Physics, Chinese Academy of Sciences,
       P.O.Box 918(4), Beijing 100039, China
       \footnote{Mailing address}}

      \date{May 18, 2001}

      \maketitle

      \begin{abstract}
      \tighten
      \indent
         In this work, we give a detailed discussion for QCD factorization
         involved in the complete chirally enhanced power corrections
       in the heavy quark limit for $B$ decays to two light pseudoscalar
       mesons, and present some detailed calculations
       of radiative corrections at the order of $\alpha_s$. We point
       out that the infrared finiteness of the vertex corrections
       in the chirally enhanced power corrections requires
       twist-3 light-cone distribution amplitudes (LCDAs) of the light
       pseudoscalar symmetric. However, even in the symmetric condition,
       there is also a logarithmic divergence from the endpoints of the
       twist-3 LCDAs in the hard spectator scattering. We point out that
       the decay amplitudes of $B\to PP$ predicted by QCD factorization
       are really free of the renormalization scale dependence, at least
       at the order of $\alpha_s$.  At last, we briefly compare the QCD
       factorization with the generalized factorization and PQCD method.
       \end{abstract}


      \vspace{1.5cm}

      {\bf PACS numbers 13.25.Hw 12.38.Bx}

      \newpage

      \narrowtext 
      \tighten

      \section{introduction}

         The study of $B$ decays plays an important role in understanding the
      origin of $CP$ violation and physics of heavy flavor. We expect that the
      parameters of the Cabibbo-Kobayashi-Maskawa (CKM) matrix in the
      standard model, for instance, the three angles $\alpha$,
      $\beta$ and $\gamma$ in the unitary triangle, can be well-determined
      from $B$ decays, especially from the charmless non-leptonic two-body B
      decays. Experimentally, many $B$ experiment projects have been running
      (CLEO, BaBar, Belle etc.), or will run in forthcoming years (BTeV,
      CERN LHCb, DESY HeraB etc.). With the accumulation of the data,
      the theorists will be urged to gain a deeper sight into $B$ decays,
      and to reduce the theoretical errors in determining the CKM
      parameters from the experimental data.

      In the theoretical frame, the standard approach to deal with such
      decays is based on the low-energy effective Hamiltonian which is
      obtained by the Wilson operator product expansion method (OPE).  In
      this effective Hamiltonian, the short-distance contributions from
      the scale above $\mu \simeq m_b$ have been absorbed into the Wilson
      coefficients with the perturbative theory and renormalization group
      method. The Wilson coefficients have been evaluated to
      next-to-leading order. Then the main task in studying non-leptonic
      two-body $B$ decays is to calculate the
      hadronic matrix elements of the effective operators. However, we do not
      have a reliable approach to evaluate them from the first principles
      of QCD dynamics up to now.

      Generally, we must resort to the factorization assumption to
      calculate the hadronic matrix elements for non-leptonic $B$ decays, in
      which the hadronic matrix element of the effective operator (in
      general, which is in the form of current-current four-quark
      operator) can be approximated as a product of two single current
      hadronic matrix elements; then it is
      parameterized into meson decay constant and meson-meson
      transition form factor. The most popular factorization model is
      the Bauer-Stech-Wirbel (BSW) model\cite{BSW}. In many cases, BSW
      model achieves great success, which can predict the branching ratios
      of many modes of non-leptonic $B$ decays in correct order of
      magnitude. This factorization assumption does hold in the limit that
      the soft interactions in the initial and final states can be
      ignored. It seems that the argument
      of color-transparency can give reasonable support to the above limit.
      Because $b$ quark is heavy, the quarks from $b$ quark decay move so fast that
      a pair of quarks in a small color-singlet object decouple from the soft
      interactions. But the shortcomings of this simple model are obvious.
      First, the renormalization scheme and scale dependence in the hadronic
      matrix elements of the effective operators are apparently missed. Then the
      full decay amplitude predicted by BSW model remains dependent on the
       renormalization scheme and scale, which are mainly from
      Wilson coefficients. In past years, many researchers improved the simple
      factorization scheme and made many remarkable progresses, such as scheme
      and scale independent effective Wilson coefficients\cite{ali1,chy2},
      effective color number which is introduced to compensate the
      `non-factorizable' contributions, etc. Furthermore, some progresses
      in nonperturbative methods, such as
      lattice QCD, QCD sum rule etc.\cite{lattice,pball,ruckle},
      allow us to compute many
      non-perturbative parameters in $B$ decays, such as the meson decay
      constants and meson-meson transition form factors. Every improvement
      allows us to have a closer look at the $B$ nonleptonic decays.

      Except for the factorization approximation, another
      important approach has been applied to study many $B$ exclusive hadronic
      decay channels, such as $B \to D\pi$, $\pi\pi$, $\pi K$ etc.
      This is PQCD method\cite{brodsky1,lihn,lucd}. In this method, people
      assumes that $B$ exclusive hadronic decay is dominated by hard gluon
      exchange. It is analogous to the framework of perturbative factorization
      for exclusive processes in QCD at large momentum transfer, such as
      the calculation of the electromagnetic form factor of the pion
      \cite{brodsky}.
      The decay amplitude for $B$ decay can be written
      as a convolution of a hard-scattering kernel with light-cone
      wave functions of the participating mesons.
         Furthermore,
       in Ref.\cite{lihn,lucd} the Sudakov
      suppression has been taken into account.

       Two year ago, Beneke, Buchalla, Neubert, and Sachrajda (BBNS) gave
      a QCD factorization formula in the heavy quark limit for the decays
      $B \to \pi \pi$\cite{beneke}. They pointed out that the radiative
      corrections from hard gluon exchange can be calculated by use of the
      perturbative QCD method if one neglects the power contributions of
      $\Lambda_{QCD}/m_b$.  This factorization formula can
      be justified in case that the ejected meson from the $b$ quark decay is
      a light meson or an onium, no matter whether the other recoiling
      meson which absorbs the spectator quark in $B$ meson is light or
      heavy. But for the case that the ejected meson is
       in an extremely asymmetric configuration, such as D meson,
       this factorization formula does not hold.
      The contributions from the hard scattering with the spectator quark in B
      meson are also involved in their formula. This kind of contribution
      cannot be contained in the naive factorization. But it appears in
      the order of $\alpha_s$. So they said that the naive factorization
      can be recovered if one neglects the radiative corrections and power
      $\Lambda_{QCD}/m_b$ suppressed contributions in the QCD factorization,
      and the `non-factorizable' contributions in the naive factorization
      can be calculated perturbatively, then we do not need a
      phenomenological parameter $N_c^{eff}$ to compensate the
      `non-factorizable' effects any more\cite{chy1,ali,chy3}.

         This QCD factorization (BBNS approach) has been applied to
      study many $B$ meson decay modes, such as $B \to D^{(*)}
      \pi^-$\cite{chay,nucl}, $\pi \pi$, $\pi K$\cite{our,ymz,osaka}
      and other interesting channels\cite{ymz1,hxg,chay1,chy}.
       Some theoretical generalizations of BBNS approach
      have also been made, such as the chirally enhanced power
      corrections\cite{ymz,osaka,our1,beneke1}
      from the twist-3 light-cone distribution amplitudes of the light
      pseudoscalar mesons. In this work, we will take a closer look at
      this issue. This work is organized as follows: Sect. II is devoted
      to a sketch of the low energy effective Hamiltonian; in Sect.III, we
      will give a detailed overview of QCD factorization, in which some
      elaborate calculations are shown, especially for the chirally
      enhanced power corrections; Sect. IV is for some detailed discussions and
      comparison of BBNS approach to the generalized factorization and
      PQCD method; we conclude in Sect.V with a summary.


      \section{Effective Hamiltonian --- First Step Factorization}

      $B$ decays involve three characteristic scales which are strongly
      ordered: $m_W \gg m_b \gg \Lambda_{QCD}$. How to separate or
      factorize these three scales is the most essential question in B
      hadronic decays.

        With the operator product expansion method (OPE),
       the relevant $\vert\Delta B\vert=1$ effective Hamiltonian
       is given by \cite{buras}:

     \begin{eqnarray}
     {\cal{H}}_{eff}&=& \frac{G_F}{\sqrt{2}}
     \left [ \sum_{q=u,c} v_q \left( C_1(\mu) Q^q_1(\mu)+ C_2(\mu)Q^q_2(\mu)
     + \sum_{k=3}^{10} C_k(\mu)Q_k(\mu)  \right) \right. \nonumber \\
      && - \left. v_t\left(
     C_{7\gamma}(\mu)Q_{7\gamma}(\mu)+C_{8G}(\mu)Q_{8G}(\mu)
      \right)\right ]+h.c.,
     \end{eqnarray}

      where
      $v_q=V_{qb}V_{qd}^{*}$(for $b\to d$ transition) or
      $v_q=V_{qb}V_{qs}^{*}$(for $b\to s$ transition)
      and $C_i(\mu)$ are the Wilson coefficients which have been evaluated to
      next-to-leading order approximation with
      the perturbative theory and renormalization group method.

       In the Eq.(1), the four-quark operators $Q_i$ are given by
      \begin{equation}
      \begin{array}{l}
      \begin{array}{ll}
      Q^u_1= ( \bar{u}_{\alpha} b_{\alpha} )_{V-A}
               ( \bar{q}_{\beta} u_{\beta} )_{V-A}&
      Q^c_1= ( \bar{c}_{\alpha} b_{\alpha} )_{V-A}
               ( \bar{q}_{\beta} c_{\beta} )_{V-A}\\
      Q^u_2= ( \bar{u}_{\alpha} b_{\beta} )_{V-A}
               ( \bar{q}_{\beta} u_{\alpha} )_{V-A}&
      Q^c_2= ( \bar{c}_{\alpha} b_{\beta} )_{V-A}
               ( \bar{q}_{\beta} c_{\alpha} )_{V-A}\\
      Q_3= (\bar{q}_{\alpha} b_{\alpha} )_{V-A}
            \sum\limits_{q'}
           ( \bar{q}^{'}_{\beta} q^{'}_{\beta} )_{V-A}&
      Q_4= (\bar{q}_{\beta} b_{\alpha} )_{V-A}
            \sum\limits_{q'}
           ( \bar{q}^{'}_{\alpha} q^{'}_{\beta} )_{V-A}\\
      Q_5= (\bar{q}_{\alpha} b_{\alpha} )_{V-A}
            \sum\limits_{q'}
            ( \bar{q}^{'}_{\beta} q^{'}_{\beta} )_{V+A}&
      Q_6= (\bar{q}_{\beta} b_{\alpha} )_{V-A}
            \sum\limits_{q'}
           ( \bar{q}^{'}_{\alpha} q^{'}_{\beta} )_{V+A}\\
      Q_7= \frac{3}{2} (\bar{q}_{\alpha} b_{\alpha} )_{V-A}
            \sum\limits_{q'} e_{q'}
           ( \bar{q}^{'}_{\beta} q^{'}_{\beta} )_{V+A}&
      Q_8=\frac{3}{2}  (\bar{q}_{\beta} b_{\alpha} )_{V-A}
         \sum\limits_{q'} e_{q'}
          ( \bar{q}^{'}_{\alpha} q^{'}_{\beta} )_{V+A}\\
      Q_9= \frac{3}{2} (\bar{q}_{\alpha} b_{\alpha} )_{V-A}
            \sum\limits_{q'} e_{q'}
          ( \bar{q}^{'}_{\beta} q^{'}_{\beta} )_{V-A}&
      Q_{10}=\frac{3}{2}  (\bar{q}_{\beta} b_{\alpha} )_{V-A}
            \sum\limits_{q'} e_{q'}
           ( \bar{q}^{'}_{\alpha} q^{'}_{\beta})_{V-A}\\
      \end{array} \\

      \end{array}
      \end{equation}
      and
      \begin{equation}
      Q_{7\gamma}=\frac{e}{8\pi^2} m_b \bar{q}_{\alpha} \sigma^{\mu\nu}
      (1+\gamma_5) b_{\alpha} F_{\mu\nu}, ~~
      Q_{8G}=\frac{g}{8\pi^2} m_b \bar{q}_{\alpha} \sigma^{\mu\nu}
      t^{a}_{\alpha \beta} (1+\gamma_5) b_{\beta} G^a_{\mu\nu}, ~~(q=d~
      {\rm or} ~s).
      \end{equation}
      with $Q^q_1$ and $Q^q_2$ being the tree operators, $Q_3-Q_6$ the QCD
      penguin operators, $Q_7-Q_{10}$ the electroweak penguin
      operators, and $Q_{7\gamma}$, $Q_{8G}$ the magnetic-penguin
      operators.

      In this effective Hamiltonian
      for $B$ decays, the contributions from large virtual momenta of the loop
      corrections from scale $\mu={\cal O}(m_b)$ to $m_W$ are
      attributed to the Wilson coefficients, and the low energy contributions
      are fully incorporated into the matrix elements of the
      operators\cite{buras}. So the derivation of the effective
      Hamiltonian can be called ``the first step factorization''.

      To evaluate the Wilson coefficients, we must extract them at
      a large renormalization scale (for example $\mu={\cal O}(m_W)$
      in the standard model) by matching the amplitude of the
      effective Hamiltonian ($A_{eff}$) to that of the full
      theory ($A_{full}$), then evolve them by the
      renormalization group equations from the scale $\mu={\cal O}(m_W)$
      to the scale $\mu={\cal O}(m_b)$. It should be noted that
      the extraction of the Wilson coefficients $C_i$ by matching does not
      depend on the choice of the external states, if we regularize the infrared
      (and mass) singularities properly\cite{buras}. All dependence on the
      choice of external states only appears in the matrix
      elements $\langle Q_i \rangle$, and is not contained in $C_i$.
      So $C_i$ only contains the short-distance contributions from the region
      where the perturbative theory can be applied. But for the
      matrix elements $\langle Q_i \rangle$, the long-distance
      contributions appear, and are process-dependent.

         Several years ago, the perturbative corrections to the Wilson
      coefficients in SM have been evaluated to next-to-leading order with
      renormalization group method\cite{buras}. As we know, the Wilson
      coefficients are generally renormalization scheme and scale dependent.
      So, in order to cancel such dependence, we must calculate the
      hadronic matrix elements of the effective operators to the
      corresponding perturbative order with the same
      renormalization scheme and at the same scale, then we can obtain a
      complete decay amplitude which is free from those unphysical
      dependences.


      \section{QCD Factorization For $B \to PP$}

          After ``the first step factorization'', the
          decay amplitude for $B \to h_1 h_2$ can be
          written as
          \begin{equation}
          {\cal A} (B \to h_1 h_2) = \sum \limits_{i} v_i C_i(\mu)
          \langle h_1 h_2 \vert Q_i(\mu) \vert $B$ \rangle ~,
          \end{equation}
          in which, as mentioned in the previous section, the contributions
          from the large scale $m_W$ down to $m_b$ has been
          separated into the Wilson coefficients $C_i(\mu)$. The remaining
          task is to calculate the hadronic matrix elements of the
          effective operators. But for the complexity of QCD
          dynamics, it is difficult to calculate these matrix
          elements reliably from first principles. The most
          popular approximation is factorization hypothesis, in
          which the matrix element of the current-current
          operator is approximated to a product of two matrix elements of
          single current operator:
          \begin{equation}
          \langle h_1 h_2 \vert Q_i \vert $B$ \rangle \simeq \langle
          h_2 \vert J_2 \vert 0 \rangle  \langle h_1 \vert J_1
          \vert $B$ \rangle ~.
          \end{equation}
      Obviously, under this approximation, the original hadronic matrix element
      $\langle Q_i(\mu) \rangle$
      misses the dependence of the renormalization scheme and
      scale which should be used to cancel the corresponding
      dependence in the Wilson coefficients $C_i(\mu)$. A
      plausible solution to recover this scale and scheme dependence
      of $\langle Q_i \rangle$ is to calculate the radiative
      corrections. In one-loop level, they can be written as
      \cite{ali,chy3,fleischer}:
      \begin{equation}
      \langle {\bf Q} \rangle = [\hat {\bf 1} + \frac{\alpha_s}{4
      \pi} \hat{\bf m}_s +\frac{\alpha_{em}}{4 \pi} \hat{\bf
      m}_e]\cdot \langle {\bf Q} \rangle_{\rm Tree} .
      \end{equation}
      Here $\hat{\bf m}_s$ and $\hat{\bf
      m}_e$ represent the one loop corrections of QCD and QED
      respectively. Then one takes
      \begin{equation}
      \langle h_1 h_2 \vert Q_i \vert $B$ \rangle_{\rm Tree} \simeq \langle
          h_2 \vert J_2 \vert 0 \rangle \langle h_1 \vert J_1
          \vert $B$ \rangle ~.
      \end{equation}
      Therefore, the scheme and scale dependence of $\langle Q_i
      \rangle$ which are expressed in the form of $\hat{\bf m}_s$ and
      $\hat{\bf m}_e$ is recovered.
      But in quark level, $\hat{\bf m}_s$ and $\hat{\bf m}_e$ usually contain
      infrared divergences if we take the external quarks
      on-shell\cite{buras1}. To remove or regularize the infrared
      divergence, the conventional treatment is to assume that
      external quarks are off-shell by $-p^2$. But this
      introduction of the infrared cutoff $-p^2$ results in a
      gauge dependence of one-loop corrections. So how to
      factorize the infrared part of the matrix elements is a very
      subtle question. But maybe this question would get
      an important simplification in the case that the final states of B
      meson decay are two light mesons.

      Two years ago, Beneke, Buchalla, Neubert and Sachrajda
      proposed a promising QCD factorization method for $B
      \to \pi \pi$. They pointed out that in the heavy
      quark limit $m_b\gg\Lambda_{QCD}$, the hadronic matrix
      elements for $B \to \pi \pi$ can be written in the
      form
      \begin{equation}
      \langle \pi \pi \vert Q \vert $B$ \rangle = \langle \pi \vert
      J_2 \vert 0 \rangle  \langle \pi \vert J_1 \vert B
      \rangle \cdot [1+ \sum r_n \alpha_s^n +{\cal
      O}(\Lambda_{QCD}/m_b)].
      \end{equation}
      Obviously, the above formula reduces to the naive factorization
      if we neglect the power corrections in
      $\Lambda_{QCD}/m_b$ and the radiative corrections in
      $\alpha_s$. They find that the radiative corrections, which are
      dominated by hard gluon exchange, can be
      calculated systematically with the perturbative theory in the limit
      $m_b \to \infty$, in terms of the convolution of the hard scattering
      kernel and the light-cone distribution
      amplitudes of the mesons. This is also similar to the framework of
      perturbative factorization
      for exclusive processes in QCD at large momentum transfer, such as
      the calculation of the electromagnetic form factor of the pion
      \cite{brodsky}. Then a factorization formula for $B \to \pi \pi$ can
      be written as \cite{beneke}:
     \begin{equation}
     \langle \pi(p^{\prime}) \pi(q) \vert Q_i \vert B(p) \rangle =
     F^{B \to \pi}(q^2) \int_0^1 dx T^I_i(x) \Phi_{\pi}(x)
     +\int_0^1 d\xi dx dy T^{II}(\xi,x,y) \Phi_B(\xi)
     \Phi_{\pi}(x) \Phi_{\pi}(y).
     \end{equation}
     We call this factorization formalism as QCD factorization or the BBNS
     approach. In the above formula, $\Phi_B(\xi)$ and $\Phi_{\pi}(x)$ are
    the leading-twist wave functions of $B$ and pion mesons
    respectively, and the $T^{I,II}_i$ denote hard-scattering kernels which are
    calculable in perturbative theory. At the order of $\alpha_s$, the
    hard kernels $T^{I,II}$ can be depicted by Fig.1. Figures 1(a)-1(d)
    represent vertex corrections, Figs 1(e) and 1(f) penguin
      corrections, and Figs 1(g) and 1(h) hard spectator scattering.

     In the heavy quark limit, both pions are energetic.
      The pion ejected from $b$ quark decay moves so fast that it can be
     described by its leading-twist light-cone distribution amplitude. The
     $q\bar{q}$ pair
     in the ejected pion is produced as a small-size color dipole. Consequently,
      the ejected pion decouples
      from the soft gluons at leading order of $\Lambda_{QCD}/m_b$.
      Of course, only the cancellation of
      soft gluons is not enough to make the factorization hold,
      it is necessary that the $q\bar{q}$ pair also decouples from the
      collinear gluons. Both the cancellations of soft gluons and
      collinear gluons guarantee that the hard kernel $T^{I}_i$
      is of infrared finiteness. Contrast to the pion ejected from b
      quark weak decay, the recoiling pion which picks up the spectator in B
      meson can not be described by its leading-twist light-cone
      distribution amplitude(LCDA), because the spectator is transferred
      to the recoiling pion as a soft quark. Here Beneke {\it et al.} take
      the point of view that the form factor $F^{B\to \pi}$ cannot be
      calculated perturbatively. If we attempt to calculate the form
      factor within the perturbative framework, by the naive power
      counting, we find that the leading twist LCDA of pion
      does not fall fast enough to suppress the singularity at the
      endpoint where the quark from $b$ decay carries almost
      all momentum of the pion. It indicates that
      the contributions to form factor are dominated
       by the soft gluon exchange\cite{nucl}. This point of view can be
      justified also from the calculation of the form factor $F^{B\to
      \pi}$ by using light-cone sum rule (LCSR)\cite{pball,ruckle}, in
      which the dominated contribution to $F^{B\to \pi}$ comes from
      the region where the the spectator quark is transferred as a soft
      quark to the pion. So the transition form factor
      survives in the factorization formula as a nonperturbative
      parameter. However, when the spectator quark in $B$ meson interacts
      with a hard gluon from the ejected pion, the recoiling pion
      can be also described by its light-cone distribution amplitude.
      This hard spectator scattering is missed in the naive
      factorization, but calculable in the perturbative QCD at the
      leading power in $\Lambda_{QCD}/m_b$.  So with this factorization
      formula, the remaining hard part of the hadronic matrix element
      $\langle Q_i \rangle$ from the scale about $m_b$ has been factorized
      into the hard scattering kernel, and the long distance contributions
      are absorbed into the transition form factors and the
      light-cone wave functions of the participating mesons. Thus
      this is the ``final factorization'' for the two-body nonleptonic
      charmless $B$ decays.

      An explicitly technical demonstration of the above argument
      has been presented in one-loop level in Refs.\cite{beneke,nucl}.
      For $B\to D\pi$, this QCD factorization has been
      proved to two-loop order\cite{nucl}. In the literature,
      the ejected pion is represented by its leading twist
      light-cone distribution amplitude(LCDA). However, since the mass of
      $b$ quark is not asymptotically large, in particular, some power
      corrections might be enhanced by certain factors, such as the scale
      of chiral symmetry breaking $\mu_{\pi}=m_{\pi}^2/(m_u+m_d)\sim
      1.5$ GeV, and have significant effects in
      studying $B$ two-body nonleptonic charmless decays. So, in
      this manner, the chirally enhanced power corrections must be
      taken into account. Accordingly, describing the ejected
      pion by its leading twist LCDA is not enough, the
      two-particle twist-3 LCDAs must be taken into account. Below, we
      will show the elaborate results of QCD factorization in these two
      cases. For illustration, we take $\bar{B}_d^0 \to
      \pi^+\pi^-$ as an example, but the result is easily
      generalized to the cases that the final states are the other
     light pseudoscalars.


      \subsection{Leading-twist Distribution Amplitude Insertion}
       When inserting leading-twist LCDA of the light
       pseudoscalar, in the heavy quark limit, the quark constituents of
      the ejected pion can be treated as a pair of collinear massless
       quark and antiquark with the momentum $uq$ and $\bar{u}q$
       respectively ($q$ is the momentum of the ejected pion and we take
      $q$ as a hard light-cone momentum in calculation, $\bar{u}=1-u$),
       because that the contributions from the transverse momenta of the
       quarks in ejected pion are power suppressed \cite{nucl}.

      \subsubsection{Vertex Corrections}
      Now we move on to the explicit one-loop calculation of the
      diagram Figs. 1(a)-1(d) for $B\to \pi \pi$.
      For illustration, we write down the one-gluon exchange
      contribution to the $\bar{B}_d^0\to \pi^+ \pi^-$
      matrix element of the operator $Q^u_2=(\bar{u}_{\alpha} b_{\beta} )_{V-A}
      ( \bar{d}_{\beta} u_{\alpha} )_{V-A}=(\bar{d}_{\alpha} b_{\alpha} )_{V-A}
      ( \bar{u}_{\beta} u_{\beta} )_{V-A}$.

      \begin{eqnarray}
      \langle Q_2^u \rangle_{(a)}&=&-g_s^2\frac{f_{\pi}}{4} \frac{C_F}{N}
      \int_0^1 du \phi(u) \int \frac{d^4k}{(2\pi)^4}
      \frac{1}{k^2(uq-k)^2[(p-k)^2-m_b^2]} \nonumber \\
      &&\times \langle \pi^+ \vert \bar{u}_i
      \gamma^{\mu} (1-\gamma_5)\slash{\hskip -2mm} q \gamma_5
      \gamma^{\alpha} (u\slash{\hskip -2mm} q-\slash{\hskip -2.5mm}
      k)\gamma_{\mu}(1-\gamma_5) (\slash{\hskip -2.5mm}
      p-\slash{\hskip -2.5mm}k+m_b)
      \gamma_{\alpha} b_i \vert \bar{B}_d^0\rangle,
      \\
      \langle Q_2^u\rangle_{(b)}&=&g_s^2\frac{f_{\pi}}{4} \frac{C_F}{N}
      \int_0^1 du \phi(u) \int \frac{d^4k}{(2\pi)^4}
      \frac{1}{k^2(\bar{u}q-k)^2[(p-k)^2-m_b^2]} \nonumber \\
      &&\times \langle \pi^+ \vert\bar{u}_i
      \gamma^{\mu} (1-\gamma_5)(\bar{u}\slash{\hskip -2mm} q-
      \slash{\hskip -2.5mm}k)\gamma^{\alpha}\slash{\hskip -2mm} q \gamma_5
      \gamma_{\mu}(1-\gamma_5) (\slash{\hskip -2.5mm}
      p-\slash{\hskip -2.5mm}k+m_b)
      \gamma_{\alpha} b_i \vert \bar{B}_d^0\rangle,
      \\
      \langle Q_2^u \rangle_{(c)}&=&-g_s^2\frac{f_{\pi}}{4} \frac{C_F}{N}
      \int_0^1 du \phi(u) \int \frac{d^4k}{(2\pi)^4}
      \frac{1}{k^2(uq+k)^2(p-q-k)^2} \nonumber \\
      &&\times \langle \pi^+ \vert\bar{u}_i \gamma_{\alpha}
      (\slash{\hskip -2.5mm} p-
      \slash{\hskip -2mm} q-\slash{\hskip -2.5mm} k)
      \gamma^{\mu} (1-\gamma_5)\slash{\hskip -2mm} q \gamma_5
      \gamma_{\alpha} (u \slash{\hskip -2mm} q +\slash{\hskip
      -2.5mm}k) \gamma_{\mu} (1-\gamma_5) b_i \vert \bar{B}_d^0\rangle,
      \\
      \langle Q_2^u \rangle_{(d)}&=&g_s^2\frac{f_{\pi}}{4} \frac{C_F}{N}
      \int_0^1 du \phi(u) \int \frac{d^4k}{(2\pi)^4}
      \frac{1}{k^2(\bar{u}q+k)^2(p-q-k)^2} \nonumber \\
      &&\times \langle \pi^+ \vert\bar{u}_i
      \gamma_{\alpha}(\slash{\hskip -2.5mm} p-
      \slash{\hskip -2mm} q-\slash{\hskip -2.5mm} k)
      \gamma^{\mu} (1-\gamma_5) (\bar{u}\slash{\hskip -2mm}q+
      \slash{\hskip -2.5mm}k) \gamma_{\alpha}
      \slash{\hskip -2mm} q \gamma_5
      \gamma_{\mu} (1-\gamma_5) b_i \vert \bar{B}_d^0\rangle,
      \end{eqnarray}

      When we calculate the vertex corrections in the leading
      power of $\Lambda_{QCD}/m_b$, not only ultraviolet
      divergence emerges but infrared divergence does also.
      Infrared divergence arises from two regions where
      the virtuality of the loop $k$ is soft or collinear to the momentum
      of the pions. In Ref.\cite{nucl}, the authors gave an explicit
      cancellation of soft and collinear divergence in vertex corrections
      for $B\to D \pi$ in eikonal approximation. Figures 1(a),1(b)
      and 1(c),1(d) cancel the soft divergence; 1(a),1(c) and 1(b),1(d)
      cancel the collinear divergence.  For
      $B\to \pi \pi$, the cancellation is similar except
      that the collinear divergence also arises from the region where
      $k$ is collinear to the momentum of the recoiling pion.
      So Figs. 1(c),1(d) cancel not only part of soft divergence but also
      part of collinear divergence.
      Below, we give an explicit calculation of the Feynman
      diagrams Figs. 1(a)-1(d) to show the cancellation of the
      infrared divergences. In order to regularize the infrared
      divergence, there are two choices for us. One is the
      dimensional regularization (DR) scheme, in which the infrared divergence
      can be regularized into the pole terms $1/(d-4)$. In contrast to the
      dimensional regularization of ultraviolet divergence, the infrared
      divergence arises when $d\leq4$, instead of $d\geq4$
      in the case of the ultraviolet divergence. So
      the dimension $d$ in regularization for infrared divergence
      must be set to be greater than 4. This is a subtle point,
      but it will not cause any ambiguity in our calculation
      because the infrared part and ultraviolet part can be safely
      separated. The other method to
      regularize the infrared divergence is the well-known massive gluon
      (MG) scheme, in which the infrared divergence is handled by
      replacing $1/k^2$ by $1/(k^2-m_g^2)$ in the gluon propagator.
      Similar scheme has been applied in earlier computation of the radiative
      corrections for $\mu^- \to e^- \bar{\nu}_e \nu_{\mu}$, in which
      the massless photon is replaced by a massive photon. In addition,
      in our latter calculation, there are also several schemes in treating
      $\gamma_5$, the most popular two are the naive dimensional regularization
      (NDR) scheme and the 't Hooft-Veltman renormalization (HV)
       scheme. Both have been applied to calculate the Wilson coefficients
       \cite{buras}. In this work, if there is no specification,
      the NDR scheme is always applied in our calculations
        for its simplicity
       \footnote{Such choice does cause a scheme dependence
      in the matrix elements. However,
      when we choose the Wilson coefficients in the same
      scheme as for the matrix elements, the final full decay amplitude is
       free of scheme-dependence.}.

      After a straightforward calculation in DR scheme and using the
      corresponding Feynman parameter integrals listed in Appendix C,
      we obtain

     \begin{eqnarray}
     \langle Q_2^u \rangle_{(a)}&=&\frac{\alpha_s}{4\pi}\frac{C_F}{N}\langle \pi^- \vert
     \bar{d}_{\alpha}\gamma^{\mu}(1-\gamma_5)u_{\alpha}\vert 0\rangle
     \langle \pi^+ \vert \bar{u}_{\beta}\gamma_{\mu}(1-\gamma_5)b_{\beta}\vert
     \bar{B}_d^0 \rangle \nonumber \\
     &\times& \int_0^1 du \phi(u) \left\{
     \left[\frac{1}{\epsilon}-\gamma_E+\ln 4\pi+2\ln \frac{\mu}{m_b}+1+
     \frac{u}{1-u} \ln u\right] \right. \nonumber \\
     &-& \left. \frac{\Gamma(1-a)}{(4 \pi)^a}
     \left(\frac{m_b}{\mu}\right)^{2a}
     \left[ \frac{1}{a^2}+\frac{2(\ln u -1)}{a}+\ln^2 u-
     2{\rm Li}_2(1-\frac{1}{u})
     -  4\ln u+5+\frac{2\ln u}{1-u}\right] \right\}, \\
     \langle Q_2^u \rangle_{(b)}&=&\frac{\alpha_s}{4\pi}\frac{C_F}{N}\langle \pi^- \vert
     \bar{d}_{\alpha}\gamma^{\mu}(1-\gamma_5)u_{\alpha}\vert 0\rangle
     \langle \pi^+ \vert \bar{u}_{\beta}\gamma_{\mu}(1-\gamma_5)b_{\beta}\vert
     \bar{B}_d^0 \rangle \nonumber \\
     &\times& \int_0^1 du \phi(u) \left\{
     -4 \left[\frac{1}{\epsilon}-\gamma_E+\ln 4\pi+2\ln \frac{\mu}{m_b}+
     \frac{11}{4}+
     \frac{\bar{u}}{1-\bar{u}} \ln \bar{u}\right] \right. \nonumber \\
     &+& \left. \frac{\Gamma(1-a)}{(4 \pi)^a}
     \left(\frac{m_b}{\mu}\right)^{2a}
     \left[ \frac{1}{a^2}+\frac{2(\ln \bar{u} -1)}{a}+\ln^2 \bar{u}-
     2{\rm Li}_2(1-\frac{1}{\bar{u}})
     -  4\ln\bar{u}+6+\frac{2\ln \bar{u}}{1-\bar{u}}\right] \right\},\\
      \langle Q_2^u \rangle_{(c)}&=&\frac{\alpha_s}{4\pi}\frac{C_F}{N}\langle \pi^- \vert
     \bar{d}_{\alpha}\gamma^{\mu}(1-\gamma_5)u_{\alpha}\vert 0\rangle
     \langle \pi^+ \vert \bar{u}_{\beta}\gamma_{\mu}(1-\gamma_5)b_{\beta}\vert
     \bar{B}_d^0 \rangle \nonumber \\
     &\times& \int_0^1 du \phi(u) \left\{
     -4\left[\frac{1}{\epsilon}-\gamma_E+\ln 4\pi+2\ln \frac{\mu}{m_b}+
     \frac{11}{4}-\ln(-u)\right] \right. \nonumber \\
     &+& \left. \frac{\Gamma(1-a)}{(4 \pi)^a}
     \left(\frac{m_b}{\mu}\right)^{2a}
     \left[ \frac{2}{a^2}+\frac{2(\ln (-u) -2)}{a}+
     10-\frac{\pi^2}{3}-4\ln(-u)+\ln^2(-u)\right] \right\}, \\
     \langle Q_2^u \rangle_{(d)}&=&\frac{\alpha_s}{4\pi}\frac{C_F}{N}\langle \pi^- \vert
     \bar{d}_{\alpha}\gamma^{\mu}(1-\gamma_5)u_{\alpha}\vert 0\rangle
     \langle \pi^+ \vert \bar{u}_{\beta}\gamma_{\mu}(1-\gamma_5)b_{\beta}\vert
     \bar{B}_d^0 \rangle \nonumber \\
     &\times& \int_0^1 du \phi(u) \left\{
     \left[\frac{1}{\epsilon}-\gamma_E+\ln 4\pi+2\ln \frac{\mu}{m_b}+
     1-\ln(-\bar{u})\right] \right. \nonumber \\
     &-& \left. \frac{\Gamma(1-a)}{(4 \pi)^a}
     \left(\frac{m_b}{\mu}\right)^{2a}
     \left[ \frac{2}{a^2}+\frac{2(\ln (-\bar{u}) -2)}{a}+
     10-\frac{\pi^2}{3}-4\ln(-\bar{u})+\ln^2(-\bar{u})\right]
     \right\}.
    \end{eqnarray}

     In above, we have set $d=4+2a$ ($a>0$) in regularizing the
     infrared divergence. Then, after summing over all four
     diagrams, we find that all pole terms in $1/a$ are really cancelled
     before we integrate over the momentum fraction variable $u$.
     So after modified minimal subtraction ($\overline{MS}$),
     we get

    \begin{eqnarray}
     &&\overline{\langle Q_2^u \rangle}_{(a)+(b)+(c)+(d)}\nonumber \\
     &=&\frac{\alpha_s}{4\pi}\frac{C_F}{N}\langle \pi^- \vert
     \bar{d}_{\alpha}\gamma^{\mu}(1-\gamma_5)u_{\alpha}\vert 0\rangle
     \langle \pi^+ \vert \bar{u}_{\beta}\gamma_{\mu}(1-\gamma_5)b_{\beta}\vert
     \bar{B}_d^0 \rangle \nonumber \\
      &\times& \int_0^1 du \phi(u) \left[
      -18-12\ln \frac{\mu}{m_b} +\frac{u}{1-u}\ln u -\frac{4\bar{u}}{1-\bar{u}}\ln
      \bar{u}+4\ln(-u)-\ln(-\bar{u}) \right. \nonumber\\
      &-& \left. \left(\ln^2 u-\ln^2\bar{u}\right)+2\left({\rm Li}_2(1-\frac{1}{u})
      -{\rm Li}_2(1-\frac{1}{\bar{u}})\right)\right. \nonumber \\
      &-&\left. \left(\frac{2\ln u}{1-u}-
      \frac{2\ln \bar{u}}{1-\bar{u}}\right)+\left(\ln^2
      (-u)-\ln^2(-\bar{u})\right)\right].
      \label{lwas}
     \end{eqnarray}
     Assuming that the light-cone distribution amplitude $\phi(u)$ is
     symmetric, then the above equation can be simplified as
     follows:
     \begin{eqnarray}
     &&\overline{\langle Q_2^u \rangle}_{(a)+(b)+(c)+(d)}\nonumber \\
     &=&\frac{\alpha_s}{4\pi}\frac{C_F}{N}\langle \pi^- \vert
     \bar{d}_{\alpha}\gamma^{\mu}(1-\gamma_5)u_{\alpha}\vert 0\rangle
     \langle \pi^+ \vert \bar{u}_{\beta}\gamma_{\mu}(1-\gamma_5)b_{\beta}\vert
     \bar{B}_d^0 \rangle \nonumber \\
      &\times& \int_0^1 du \phi(u) \left[
      -18-12\ln \frac{\mu}{m_b}+3\frac{1-2u}{1-u}\ln u
      -3i\pi\right].\label{lws}
      \end{eqnarray}

     It is easy to check that the above equations are consistent with the
     results in previous works. Actually, with the MG scheme, we get the
     same results as that by using the DR scheme.

     With Eqs.(\ref{lwas},\ref{lws}), we can compute the vertex
corrections no matter the LCDA $\phi(u)$ is symmetric or asymmetric. This
is very important in principle. For instance, when kaon is ejected from b
quark decay, we must take the contributions from the asymmetric part of
LCDA of kaon into account,  although the contributions from the asymmetric
part are very small numerically\cite{osaka}.


      \subsubsection{Penguin Corrections}
      There are two kinds of penguin corrections. One is the four
      quark operators insertion [Fig. 1(e)]; the other is the magnetic
      penguin insertion [Fig. 1(f)]. The first kind is generally called
      BSS mechanism. In generalized factorization, BSS mechanism plays
      a very important role in $CP$ violation because it is the
      unique source of strong phases. But in generalized
      factorization, the virtuality of the gluon or photon is
      ambiguous; usually one varies $k^2$ around
      $m_B^2/2$. This variation does not have an important effect on
      the branching ratios, but it does for $CP$ asymmetries.
      In QCD factorization, this ambiguity is rendered by taking the
      virtuality of the gluon as $k^2=(p-uq)^2=\bar{u}m_b^2/2$.
 When treating penguin contractions,
one should be careful that Fig. 1(e) contains two kinds of topology, which
is depicted in Fig. 2. They are equivalent in 4 dimensions according to
Fierz rearrangement. However, since penguin corrections contain ultraviolet
divergence, we must do calculations in d dimensions where these two
kinds of topology are not equivalent \cite{buras2}.
Below we give an explicit calculation of $Q_4$ or $Q_6$ penguin insertions
for $\bar{B}_d^0 \to \pi^+ \pi^-$ which belong to the second topology,
Fig. 2(b):
      \begin{eqnarray}
      \langle Q_{4,6}\rangle_{(e)}^{\rm twist-2}&=&
       \frac{f_{\pi}}{4} g_s^2 \mu^{2\epsilon}
      \frac{C_F}{N} \int_0^1 du ~\phi(u)
      \langle \pi^+\vert \bar{u}_i \gamma_{\alpha}
     \slash{\hskip -2.0mm}q \gamma_5 \gamma_{\mu} (1- \gamma_5)
      b_i \vert \bar{B}_d^0 \rangle \nonumber \\
      &\times& \left. \sum \limits_q
      \int \frac{d^dk}{(2\pi)^d}
      \frac{-{\rm Tr}[(\slash {\hskip -2.5mm}l -\slash {\hskip -2.5mm} k -m_q)
     \gamma^{\alpha} (\slash {\hskip -2.5mm}
      k+m_q) \gamma^{\mu}
      (1\mp \gamma_5)]}{[(l-k)^2-m_q^2][k^2-m_q^2]l^2} \right |_{l=p-uq}
      \nonumber \\
      &=& -2i f_{\pi}\frac{\alpha_s}{4 \pi}
      \frac{C_F}{N} \int_0^1 du \phi(u)
      \langle \pi^+\vert \bar{u}_i \gamma_{\alpha}
     \slash{\hskip -2.0mm}q \gamma_5 \gamma_{\mu} (1- \gamma_5)
      b_i \vert \bar{B}_d^0 \rangle
       \left[\frac{l_{\alpha}l_{\mu}}{l^2}-g_{\alpha
       \mu}\right]
     \nonumber \\
      &\times& \left. \sum \limits_q \left[
\frac{1}{6}(\frac{1}{\epsilon}-\gamma_E+\ln 4
\pi)
      +\int_0^1 dt~t(1-t)\ln\frac{\mu^2}{m_q^2-t(1-t)l^2-i
      \epsilon}\right]\right|_{l=p-uq}~.
      \end{eqnarray}
      After $\overline{MS}$ subtraction and using the equations of
motions, we get the finite result
\begin{eqnarray}
\overline{\langle Q_{4,6}\rangle}_{(e)}^{\rm twist-2} &=& -\frac{\alpha_s}{4\pi} \frac{C_F}{N}
\langle \pi^- \vert \bar{d}_i \gamma^{\mu} (1-\gamma_5) u_i\vert 0 \rangle
\langle \pi^+ \vert \bar{u}_j \gamma_{\mu} (1-\gamma_5) b_j \vert
\bar{B}_d^0 \rangle
\nonumber \\
&\times& \sum \limits_q \left [ \frac{4}{3} \ln \frac{\mu}{m_b}-4 \int_0^1
du ~\phi(u) \int_0^1dt~t(1-t)
\ln(s_q-t(1-t)\bar{u}-i\epsilon)\right]~,
\end{eqnarray}
where $s_q=m_q^2/m_b^2$. The first topology, Fig. 2(a), for example,
$Q_1^c$
penguin insertion for $\bar{B}_d^0 \to \pi^+ \pi ^-$, is similar
to the results of the second topology, Fig.2(b), except that there is an
extra factor $-2/3$:
\begin{eqnarray}
\overline{\langle Q_1^c \rangle}_{(e)}^{\rm twist-2} &=& -\frac{\alpha_s}{4\pi} \frac{C_F}{N}
\langle \pi^- \vert \bar{d}_i \gamma^{\mu} (1-\gamma_5) u_i\vert 0 \rangle
\langle \pi^+ \vert \bar{u}_j \gamma_{\mu} (1-\gamma_5) b_j \vert
\bar{B}_d^0 \rangle
\nonumber \\
  &\times&  \left [ -\frac{2}{3}+\frac{4}{3} \ln \frac{\mu}{m_b}-4
\int_0^1 du ~\phi(u)\int_0^1 dt~t(1-t)
\ln(s_c-t(1-t)\bar{u}-i\epsilon)\right]~.
\end{eqnarray}

     For the magnetic penguin insertion, it is the easiest calculation of
the radiative corrections. The result of $Q_{8G}$ insertion
for $\bar{B}_d^0 \to \pi^+ \pi^-$ is

       \begin{eqnarray}
        \langle Q_{8G} \rangle_{(f)}^{\rm twist-2}&=&-\frac{\alpha_s}{4 \pi}
         \frac{C_F}{N}f_{\pi} m_b
          \left. \int_0^1 du~\phi(u) \frac{1}{k^2}
          \langle \pi^+\vert \bar{u}_i \gamma^{\alpha}\sla q
         \gamma_5\sigma_{\beta\alpha}k^{\beta}
         (1+\gamma_5)b_i\vert \bar{B}_d^0 \rangle \right |_{k=p-uq}
         \nonumber \\
         &=&- \frac{\alpha_s}{4 \pi}
         \frac{C_F}{N}\int_0^1 du~\frac{2\phi(u)}{\bar{u}}
          \langle \pi^- \vert \bar{d}_i \gamma^{\mu} (1-\gamma_5)\vert 0\rangle
         \langle \pi^+ \vert \bar{u}_i \gamma_{\mu} (1-\gamma_5) \vert
         \bar{B}_d^0 \rangle
        \end{eqnarray}

      \subsubsection{Hard scattering with the spectator}

     Hard spectator scattering [Fig.1(g) and (h)] is completely missing in
the naive factorization. But in QCD factorization, it can be calculated in
the perturbative QCD, and expressed by a convolution of the hard kernel
$T^{II}$ and the LCDAs of mesons. At the
leading power of $\Lambda_{QCD}/m_b$, both of the light pseudoscalars
from the $B$ meson decay can be represented by their leading twist
LCDAs. So after a straightforward
calculation, we obtain this contribution for $\bar{B}_d^0 \to \pi^+ \pi^-$
from the operator $Q_2^u$ insertion,

\begin{eqnarray}
\langle Q_2^u \rangle_{(g)+(h)}^{\rm twist-2} &=& \frac{-i f_B f_{\pi}^2}{64} \frac{C_F}{N^2}
g_s^2 \int_0^1 d\xi ~du ~dv ~\phi_B(\xi) ~\phi(u) ~\phi(v) \nonumber \\
&\times& \left \{ \left. \frac{{\rm Tr} [(\sl p +m_B)\gamma_5
\gamma^{\alpha} \sla q_1 \gamma_5
\gamma_{\rho} (1-\gamma_5) \sla q_2 \gamma_5 \gamma_{\alpha}
\slash{\hskip -1.5mm} l_d
\gamma_{\rho}(1-\gamma_5)]}{k^2 l_d^2}
\right |^{k=\xi p-\bar{v}q_1}_{l_d=uq_2-k}
\right. \nonumber \\
&& + \left. \left. \frac{{\rm Tr} [(\sl p +m_B)\gamma_5 \gamma^{\alpha}
\sla q_1
\gamma_5 \gamma_{\rho} (1-\gamma_5)  \slash{\hskip -1.5mm} l_u
\gamma_{\alpha}
\sla q_2 \gamma_5
\gamma_{\rho}(1-\gamma_5)]}{k^2 l_u^2}
\right |^{k=\xi p-\bar{v}q_1}_{l_u=k-\bar{u}q_2}
\right\} \nonumber \\
&=& i \pi \alpha_s f_B f_{\pi}^2 \frac{C_F}{N^2}
\int_0^1 d\xi\frac{\phi_B(\xi)}{\xi}\int_0^1 ~du \frac{\phi(u)}{u} \int_0^1~dv ~
\frac{\phi(v)}{\bar{v}} ~.
\end{eqnarray}

      \subsection{Chirally enhanced corrections --- Twist-3 LCDAs
      insertion}
It has been observed that QCD factorization is demonstrated only in the strict
heavy quark limit. This means that any generalization of QCD factorization
to include or partly include power corrections of $\Lambda_{QCD}/m_b$
should redemonstrate that factorization still holds. There are a variety
of sources which may contribute to power corrections in $1/m_b$; examples
are higher twist distribution amplitudes, transverse momenta of quarks in
the light meson, annihilation diagrams, etc. Unfortunately, there is
no known systematic way to evaluate these power corrections for
exclusive decays. Moreover, factorization might break down when these
power corrections, for instance, transverse momenta effects, are considered.
This indicates that one might have to give up such an ambitious plan that
all power corrections could be, at least in principle, incorporated into
QCD factorization order by order. One might argue that power corrections
in $B$ decays are numerically unimportant because these corrections are expanded
in order of a small number $\Lambda_{QCD}/m_b \simeq 1/15$. But this is not true.
For instance, the contributions of operator $Q_6$ to decay amplitudes would
formally vanish in the strict
heavy quark limit. However it is numerically very important in
penguin-dominated $B$ rare
decays, such as interesting channels $B \to \pi K$, etc.
This is because $Q_6$ is always multiplied by a formally power suppressed
but chirally enhanced factor $r_{\chi}=2
m_{P}^2/m_b(m_1+m_2) \sim {\cal O}(1)$, where $m_1$ and $m_2$ are
current quark masses. So power suppression might probably fail at
least in this case. Therefore phenomenological applicability of QCD
factorization in $B$ rare decays requires at least a consistent inclusion
of chirally enhanced corrections.

Chirally enhanced corrections arise from the two particle twist-3
light-cone distribution
amplitudes, generally called $\phi_p(x)$ and $\phi_{\sigma}(x)$.
So when chirally enhanced corrections are concerned, the final light
mesons should be described by leading twist and twist-3 distribution
amplitudes. Then it is crucial to show that factorization really holds
when considering twist-3 distribution amplitudes. The most difficult part
is to demonstrate the infrared finiteness of the hard scattering kernels
$T^{I}_i$. In addition, possible chirally enhanced power corrections can
also appear in the hard spectator scattering. So, for consistency, we must
involve these corrections.

       \subsubsection{Vertex corrections}
 When we calculate the chirally enhanced power corrections,
 contrast to the leading-twist light-cone wave function insertion,
 the cancellation of the infrared divergences in the vertex corrections
to $(V-A)(V+A)$ operator (here it is $Q_5$ or $Q_7$)
can not be shown simply by the eikonal approximation similar to what has
been
done at the leading power of $\Lambda_{QCD}/m_b$, because the Dirac
structure or spin structure of twist-3 light-cone wave functions of the
light pseudoscalar makes the ``on-shell'' condition
for the external quarks invalid. Thus, to justify
the cancellation of the infrared divergence in $(V-A)(V+A)$ vertex
corrections, we must give the explicit calculation. As mentioned in the
previous subsections, we have two choices to regularize the infrared
divergence in one-loop calculation. One is the DR scheme; the other is MG
scheme. Generally, these two schemes are equivalent, for instance,
similar to  what has been done in $(V-A)(V-A)$ vertex corrections.
However, in
the DR scheme, it is difficult to extrapolate the twist-3 wave
functions of the light pseudoscalar to d dimensions properly, although they are
well-defined in 4 dimensions. Therefore, we prefer to use the MG
scheme in our calculation for chirally enhanced corrections to avoid the
above possible problems.

     In addition, generally we calculate the Feynman diagrams in the momentum space,
so the correct projection of the light-cone wave functions of the light
pseudoscalar in the momentum space is necessary. From Appendix B, we find that it is
easy to find the proper momentum space projection of the leading twist and $\phi_p$
type twist-3 wave function, but for $\phi_{\sigma}$, the projection is not very clear.
Note that the coordinate $x^{\mu}$ in the definition of $\phi_{\sigma}$ by the
non-local matrix element must be transformed into a partial derivative of a
certain momentum in the projection of momentum space. However, it is difficult
to find the derivative which makes the projection only depend on the structure
of the light pseudoscalar itself. Generally, the momentum which the partial
derivative acts on is dependent on the hard kernel. Therefore, we prefer to
compute the Feynman diagrams of the twist-3 wave functions insertion,
especially
$\phi_{\sigma}$ insertion in the coordinate space.
We think that such calculation can avoid the ambiguity about how to project the coordinate
$x^{\mu}$ into the momentum space. We recalculate the leading twist insertion by using
the same method, and obtain the same results as those in the previous sections.
Below, we will show how to perform this trick in calculation of
$\phi_{\sigma}$
insertion. For example, let us consider Fig. 3. In
coordinate space, we have

\begin{eqnarray}
{\rm Fig.~3} &=&\frac{f_P\mu_P}{4}\frac{C_F}{N} g_s^2 \int du~
\frac{\phi_{\sigma}(u)}{6}~
\int d^4 x_1 d^4 x_2  \int \frac{d^4 k}{(2 \pi)^4}
\frac{d^4 l_{\bar{u}}}{(2 \pi)^4} \frac{d^4 l_b}{(2 \pi)^4}
\frac{e^{i(\bar{u}q-k+l_{\bar{u}})\cdot x_2} e^{i(k+l_b-p)\cdot x_1}}
{(k^2-m_g^2) l_{\bar{u}}^2 (l_b^2-m_b^2)}
\nonumber \\
&\times&\langle \pi^+\vert{\bar u_i}~
\gamma^{\rho}(1+\gamma_5)
\slash{\hskip -1.5mm}l_{\bar{u}}
\gamma^{\alpha}\gamma_5 \sigma_{\mu \nu}
q^{\mu} x_2^{\nu}
\gamma_{\rho}(1-\gamma_5)
(\slash{\hskip -1.5mm}l_b+m_b)
\gamma_{\alpha} b_i\vert \bar{B}_d^0 \rangle~
\nonumber \\
&=&\frac{f_P\mu_P}{4}\frac{C_F}{N} g_s^2 \int du~
\frac{\phi_{\sigma}(u)}{6}~
 \int \frac{d^4 k}{(2 \pi)^4}
~\frac{d^4 l_{\bar{u}} d^4 x_2 }{(2 \pi)^4} ~~
\frac{e^{i(\bar{u}q-k+l_{\bar{u}})\cdot x_2}~q^{\mu} ~x_2^{\nu}}
{(k^2-m_g^2) l_{\bar{u}}^2 (l_b^2-m_b^2)}
\nonumber \\
 &\times&\left. \langle \pi^+\vert{\bar u_i}~
\gamma^{\rho}(1+\gamma_5)
\slash{\hskip -1.5mm}l_{\bar{u}}
\gamma^{\alpha}\gamma_5 \sigma_{\mu \nu}
\gamma_{\rho}(1-\gamma_5)
(\slash{\hskip -1.5mm}l_b+m_b)
\gamma_{\alpha} b_i\vert \bar{B}_d^0 \rangle\right|_{l_b=p-k}
\nonumber \\
&=&  \frac{i f_P\mu_P}{4}\frac{C_F}{N} g_s^2 \int du~
\frac{\phi_{\sigma}(u)}{6}~
 \int \frac{d^4 k}{(2 \pi)^4}
\frac{q^{\mu} }
{(k^2-m_g^2) (l_b^2-m_b^2)}
\nonumber \\
&\times&\left. \frac{\partial}{\partial l_{\bar{u}\nu}}\left\{
\langle \pi^+\vert{\bar u_i}~
\gamma^{\rho}(1+\gamma_5)
\frac{\slash{\hskip -1.5mm}l_{\bar{u}}}{l_{\bar{u}}^2}
\gamma^{\alpha}\gamma_5 \sigma_{\mu \nu}
\gamma_{\rho}(1-\gamma_5)
(\slash{\hskip -1.5mm}l_b+m_b)
\gamma_{\alpha} b_i\vert \bar{B}_d^0 \rangle\right\}
\right|^{l_{\bar{u}}=k-\bar{u}q}_{l_b=p-k}~.
\end{eqnarray}
The above trick has been applied in the calculation of the proper correlation
function to extract the transition form factor $F^{B\to \pi}$  within the
frame of the light-cone sum rule\cite{ruckle}.
So within the same frame, we obtain
\begin{eqnarray}
      \langle Q_5 \rangle_{(a)}&=&g_s^2\frac{f_{\pi}\mu_{\pi}}{4} \frac{C_F}{N}
      \int_0^1 du  \int \frac{d^4k}{(2\pi)^4}
      \frac{1}{(k^2-m_g^2) l_d^2(l_b^2-m_b^2)} \nonumber \\
      &&\times \left\{ \phi_p(u)\langle \pi^+ \vert \bar{u}_i
      \gamma^{\mu} (1+\gamma_5) \gamma_5
      \gamma^{\alpha} \slash{\hskip -1.5mm} l_d
      \gamma_{\mu}(1-\gamma_5) (\slash{\hskip -1.5mm}l_b+m_b)
      \gamma_{\alpha} b_i \vert \bar{B}_d^0\rangle \right. \nonumber \\
      &&+\left. i\frac{\phi_{\sigma}(u)}{6}
      \langle \pi^+ \vert \bar{u}_i
       \gamma^{\rho} (1+\gamma_5) \gamma_5 \sigma_{\mu \nu} q^{\mu}  \gamma^{\alpha}
      \left(\gamma^{\nu}-\frac{2l_d^{\nu}\slash{\hskip -1.5mm}l_d}{l_d^2}\right)
       \right.\nonumber \\
      &&  ~~~~~~~~~~~~~~~~~~~~~~~~\left. \left.
      \gamma_{\rho}(1-\gamma_5) (\slash{\hskip -1.5mm}l_b+m_b)
      \gamma_{\alpha} b_i \vert \bar{B}_d^0\rangle \right\}
      \right|^{l_d=uq-k}_{l_b=p-k},
      \\
      \langle Q_5 \rangle_{(b)}&=&g_s^2\frac{f_{\pi}\mu_{\pi}}{4} \frac{C_F}{N}
      \int_0^1 du  \int \frac{d^4k}{(2\pi)^4}
      \frac{1}{(k^2-m_g^2)l_{\bar{u}}^2(l_b^2-m_b^2)} \nonumber \\
      &&\times \left\{\phi_p(u)\langle \pi^+ \vert\bar{u}_i
      \gamma^{\mu} (1+\gamma_5)
      \slash{\hskip -1.5mm}l_{\bar{u}}\gamma^{\alpha}\gamma_5
      \gamma_{\mu}(1-\gamma_5) (\slash{\hskip -1.5mm} l_b+m_b)
      \gamma_{\alpha} b_i \vert \bar{B}_d^0\rangle \right.\nonumber \\
       &&+\left. i\frac{\phi_{\sigma}(u)}{6}
      \langle \pi^+ \vert \bar{u}_i
      \gamma^{\rho} (1+\gamma_5)
        \left(\gamma^{\nu}-\frac{2l_{\bar{u}}^{\nu}\slash{\hskip -1.5mm}l_{\bar{u}}}
      {l_{\bar{u}}^2}\right)
         \gamma^{\alpha} \gamma_5 \sigma_{\mu \nu} q^{\mu} \right.\nonumber \\
      &&  ~~~~~~~~~~~~~~~~~~~~~~~~\left. \left.
      \gamma_{\rho}(1-\gamma_5) (\slash{\hskip -1.5mm}l_b+m_b)
      \gamma_{\alpha} b_i \vert \bar{B}_d^0\rangle \right\}
      \right|^{l_{\bar{u}}=k-\bar{u}q}_{l_b=p-k},
      \\
      \langle Q_5 \rangle_{(c)}&=&g_s^2\frac{f_{\pi}\mu_{\pi}}{4} \frac{C_F}{N}
      \int_0^1 du  \int \frac{d^4k}{(2\pi)^4}
      \frac{1}{(k^2-m_g^2)l_d^2l_u^2} \nonumber \\
      &&\times \left\{\phi_p(u)\langle \pi^+ \vert\bar{u}_i \gamma_{\alpha}
      \slash{\hskip -1.5mm} l_u
      \gamma^{\mu} (1+\gamma_5)\gamma_5
      \gamma_{\alpha} \slash{\hskip -1.5mm} l_d
       \gamma_{\mu} (1-\gamma_5) b_i \vert \bar{B}_d^0\rangle \right. \nonumber \\
      &&+\left. i\frac{\phi_{\sigma}(u)}{6}
      \langle \pi^+ \vert \bar{u}_i \gamma^{\alpha}
      \slash{\hskip -1.5mm} l_u
      \gamma^{\mu} (1+\gamma_5)\gamma_5 \sigma_{\mu \nu} q^{\mu} \gamma_{\alpha}\right.
      \nonumber \\
      &&  ~~~~~~~~~~~~~~~~~~~~~~~~\left. \left.
        \left(\gamma^{\nu}-\frac{2l_d^{\nu}\slash{\hskip -1.5mm}l_d}{l_d^2}\right)
       \gamma_{\rho}(1-\gamma_5) b_i \vert \bar{B}_d^0\rangle \right\}
      \right|^{l_d=uq-k}_{l_u=p-q+k},
      \\
      \langle Q_5 \rangle_{(d)}&=&g_s^2\frac{f_{\pi}\mu_{\pi}}{4} \frac{C_F}{N}
      \int_0^1 du  \int \frac{d^4k}{(2\pi)^4}
      \frac{1}{(k^2-m_g^2) l_{\bar{u}}^2 l_u^2} \nonumber \\
      &&\times \left\{ \phi_p(u)\langle \pi^+ \vert\bar{u}_i
      \gamma^{\alpha}\slash{\hskip -1.5mm} l_u
      \gamma^{\mu} (1+\gamma_5) \slash{\hskip -1.5mm}l_{\bar{u}}
       \gamma_{\alpha} \gamma_5
      \gamma_{\mu} (1-\gamma_5) b_i \vert \bar{B}_d^0\rangle\right.\nonumber \\
       &&  +\left. i\frac{\phi_{\sigma}(u)}{6}
      \langle \pi^+ \vert \bar{u}_i
       \gamma^{\alpha}\slash{\hskip -1.5mm} l_u
      \gamma^{\rho} (1+\gamma_5)
       \left(\gamma^{\nu}-\frac{2 l_{\bar{u}}^{\nu}\slash{\hskip -1.5mm}l_{\bar{u}}}
      {l_{\bar{u}}^2}\right)
        \right. \nonumber \\
       &&  ~~~~~~~~~~~~~~~~~~~~~~~~\left. \left.
       \gamma_{\alpha}  \gamma_5 \sigma_{\mu \nu} q^{\mu}
       \gamma_{\rho}(1-\gamma_5) b_i \vert \bar{B}_d^0\rangle \right\}
      \right|^{l_{\bar{u}}=k-\bar{u}q}_{l_u=p-q+k}.
       \end{eqnarray}

       Perform the one-loop integrations:

     \begin{eqnarray}
      \langle Q_5 \rangle_{(a)}&=&2\frac{\alpha_s}{4 \pi} \frac{C_F}{N}
      \langle\pi^- \vert \bar{d}_i (1+\gamma_5) u_i\vert 0\rangle
      \langle \pi^+ \vert \bar{u}_j (1-\gamma_5) b_j \vert \bar{B}_d^0 \rangle
      \nonumber \\
      &\times&\int_0^1 du~
       \left\{\phi_p(u)\left[-(\frac{1}{\epsilon}-\gamma_E+\ln 4\pi +2\ln \frac{\mu}{m_b})
        \right.\right.\nonumber \\
       &+&\left. \left.\frac{1}{4}\ln^2 \lambda+\ln(-u)\ln\lambda-2\ln u \ln\lambda
       +\frac{1}{2}\ln^2 u-{\rm Li}_2(1-\frac{1}{u})+\frac{1}{2}+\frac{5}{4}
       \pi^2\right]\right. \nonumber \\
      &+&\left. \frac{\phi_{\sigma}(u)}{6u}
      \left [\frac{1}{2}\ln^2 \lambda + 2 \ln(-u)\ln\lambda
      -4 \ln u \ln \lambda+\ln \lambda
      \right. \right. \nonumber \\
       &+&\left. \left.4\ln^2 u-\ln^2(-u)-2\ln u \ln(1-u)+2{\rm Li}_2(\frac{1}{u})
        +\frac{7}{6}\pi^2-\ln u -\frac{\ln u}{1-u}\right]\right\},
       \\
       \langle Q_5 \rangle_{(b)}&=&-2\frac{\alpha_s}{4 \pi} \frac{C_F}{N}
      \langle\pi^- \vert \bar{d}_i (1+\gamma_5) u_i\vert 0\rangle
      \langle \pi^+ \vert \bar{u}_j (1-\gamma_5) b_j \vert \bar{B}_d^0 \rangle
      \nonumber \\
      &\times&\int_0^1 du~
       \left\{\phi_p(u)\left[-(\frac{1}{\epsilon}-\gamma_E+\ln 4 \pi +2\ln \frac{\mu}{m_b})
        \right.\right.\nonumber \\
       &+&\left. \left.\frac{1}{4}\ln^2 \lambda+\ln(-\bar{u})\ln\lambda
       -2\ln \bar{u} \ln\lambda
       +\frac{1}{2}\ln^2 \bar{u}-{\rm Li}_2(1-\frac{1}{\bar{u}})
        -\frac{5}{2}+\frac{5}{4}
       \pi^2\right]\right. \nonumber \\
       &+&\left.\frac{\phi_{\sigma}(u)}{6\bar{u}}
      \left [\frac{1}{2}\ln^2 \lambda + 2 \ln(-\bar{u})\ln\lambda-
       4 \ln \bar{u} \ln \lambda
      +\ln \lambda
      \right. \right. \nonumber \\
       &+&\left. \left. 4\ln^2 \bar{u}-\ln^2(-\bar{u})-2\ln \bar{u}
      \ln(1-\bar{u})+2{\rm Li}_2(\frac{1}{\bar{u}})
        +\frac{7}{6}\pi^2-\ln \bar{u} -\frac{\ln
      \bar{u}}{1-\bar{u}}\right]\right\},
       \\
       \langle Q_5\rangle_{(c)}&=&-2\frac{\alpha_s}{4 \pi} \frac{C_F}{N}
      \langle\pi^- \vert \bar{d}_i (1+\gamma_5) u_i\vert 0\rangle
      \langle \pi^+ \vert \bar{u}_j (1-\gamma_5) b_j \vert \bar{B}_d^0 \rangle
      \nonumber \\
       &\times&\int_0^1 du~
       \left\{\phi_p(u)\left[-(\frac{1}{\epsilon}-\gamma_E+\ln 4\pi+2\ln\frac{\mu}{m_b})
        \right. \right. \nonumber \\
      &+&\left. \left.\frac{1}{2}\ln^2\lambda-[\ln(-u)-2]\ln \lambda+
        \frac{1}{2}\ln^2(-u)-\ln(-u)-\frac{3}{2}+\frac{\pi^2}{3}\right]\right.
     \nonumber \\
     &+& \left. \frac{\phi_{\sigma}(u)}{6u}
      \left [\ln^2 \lambda - [2\ln(-u)-3]\ln\lambda
      \right. \right.\nonumber \\
       &+&\left. \left. 2\ln(-u)\ln u -\ln^2 u-3\ln(-u)
        +3-\frac{\pi^2}{3} \right] \right\},
       \\
      \langle Q_5 \rangle_{(d)}&=&2\frac{\alpha_s}{4 \pi} \frac{C_F}{N}
      \langle\pi^- \vert \bar{d}_i (1+\gamma_5) u_i\vert 0\rangle
      \langle \pi^+ \vert \bar{u}_j (1-\gamma_5) b_j \vert \bar{B}_d^0 \rangle
      \nonumber \\
       &\times&\int_0^1 du~
       \left\{\phi_p(u)\left[-(\frac{1}{\epsilon}-\gamma_E+\ln 4\pi+2\ln\frac{\mu}{m_b})
        \right. \right. \nonumber \\
      &+&\left. \left.\frac{1}{2}\ln^2\lambda-[\ln(-\bar{u})-2]\ln
\lambda+
        \frac{1}{2}\ln^2(-\bar{u})-\ln(-\bar{u})+\frac{3}{2}+\frac{\pi^2}{3}\right]\right.
     \nonumber \\
      &+&\left. \frac{\phi_{\sigma}(u)}{6\bar{u}}
      \left [\ln^2 \lambda - [2\ln(-\bar{u})-3]\ln\lambda
      \right. \right.\nonumber \\
       &+&\left. \left. 2\ln(-\bar{u})\ln \bar{u} -\ln^2 \bar{u}-3\ln(-\bar{u})
        +3-\frac{\pi^2}{3} \right]   \right\}.
       \end{eqnarray}
       Here $\lambda=m_g^2/m_b^2$.
         From the above equations, it is observed that,
in the case of $\phi_{\sigma}$ distribution amplitudes, the terms with
infrared divergence in vertex
correction diagrams cannot cancel unless $\phi_{\sigma}(u)$
is a symmetric function: $\phi_{\sigma}(u)=\phi_{\sigma}(\bar u)$.
This is an unexpected result, which means QCD factorization is violated
for asymmetric twist-3 light-cone distribution amplitudes. This indicates
that chirally enhanced corrections can be included consistently in the
framework of QCD factorization only when twist-3 light-cone distribution
amplitudes are symmetric. Therefore, in the following, we will implicitly
assume a symmetric  twist-3 light-cone distribution amplitude for light
pseudoscalar mesons. It is then straightforward to show that vertex
corrections of $(V-A)(V+A)$ operator are completely cancelled
after summing over four diagrams in the case of $\phi_{\sigma}$
distribution amplitude. The final result of $(V-A)(V+A)$ vertex corrections,
in the condition that the twist-3 LCDA is symmetric, is
    \begin{equation}
     \langle Q_5 \rangle_{(a)+(b)+(c)+(d)}=12\frac{\alpha_s}{4\pi}\frac{C_F}{N}
     \langle\pi^- \vert \bar{d}_i (1+\gamma_5) u_i\vert 0\rangle
      \langle \pi^+ \vert \bar{u}_j (1-\gamma_5) b_j \vert \bar{B}_d^0 \rangle
     \end{equation}

      \subsubsection{Penguin corrections}

     In quark level, usually one decomposes the basic QCD vertex $-ig_s\gamma^{\mu}T^a_{ij}$
in penguin insertion into the two chiral current couplings
$-i\frac{1}{2}g_s\gamma^{\mu}T^a_{ij}(1+\gamma_5)$ and
$-i\frac{1}{2}g_s\gamma^{\mu}T^a_{ij}(1-\gamma_5)$; then the
penguin insertions contribute the same magnitude to the $(V-A)(V-A)$ and $(V-A)(V+A)$
vertex. But in hadron level, this point of view must be examined in elaborate calculation.

    For illustration, we give the results of $Q_4$ or $Q_6$ penguin corrections to
$\bar{B}_d^0 \to \pi^+ \pi^-$, which belong to the second penguin topology
Fig. 2(b),
when $\phi_p(u)$ is inserted:

        \begin{eqnarray}
      \langle Q_{4,6}\rangle_{(e)}^{\phi_p}&=&- \frac{f_{\pi}\mu_{\pi}}{4} g_s^2 \mu^{2\epsilon}
      \frac{C_F}{N} \int_0^1 du ~\phi_p(u)
      \langle \pi^+\vert \bar{u}_i \gamma_{\alpha}
   \gamma_5 \gamma_{\mu} (1- \gamma_5)
      b_i \vert \bar{B}_d^0 \rangle \nonumber \\
      &\times& \sum \limits_q
      \int \frac{d^dk}{(2\pi)^d}
      \frac{-{\rm Tr}[(\slash {\hskip -2.5mm} p - u \slash {\hskip
      -2mm}q -\slash {\hskip -2.5mm} k -m_q)\gamma^{\alpha} (\slash
     {\hskip -2.5mm}
      k+m_q) \gamma^{\mu}
      (1\mp \gamma_5)]}{[(p-uq-k)^2-m_q^2][k^2-m_q^2](p-uq)^2}
      \nonumber \\
      &=& 2i f_{\pi}\mu_{\pi}\frac{\alpha_s}{4 \pi}
      \frac{C_F}{N} \int_0^1 du~ \phi_p(u)
      \langle \pi^+\vert \bar{u}_i \gamma_{\alpha}
     \gamma_5 \gamma_{\mu} (1- \gamma_5)
      b_i \vert \bar{B}_d^0 \rangle
       \left[\frac{l_{\alpha}l_{\mu}}{l^2}-g_{\alpha
       \mu}\right]
     \nonumber \\
      &\times& \left. \sum \limits_q \left[
     \frac{1}{6}(\frac{1}{\epsilon}-\gamma_E+\ln 4 \pi)
      +\int_0^1 dt~t(1-t)\ln\frac{\mu^2}{m_q^2-t(1-t)l^2-i\epsilon}
     \right]\right|_{l=p-uq} \nonumber \\
      \end{eqnarray}

      After $\overline{MS}$ subtraction, we obtain

     \begin{eqnarray}
      \overline {\langle Q_{4,6}\rangle}_{(e)}^{\phi_p}&=& 2\frac{\alpha_s}{4\pi} \frac{C_F}{N}
\langle \pi^- \vert \bar{d}_i (1+\gamma_5) u_i\vert 0 \rangle
\langle \pi^+ \vert \bar{u}_j  (1-\gamma_5) b_j \vert
\bar{B}_d^0 \rangle
\nonumber \\
&\times& \sum \limits_q \left [  \ln \frac{\mu}{m_b}-3
\int_0^1 du~\phi_p(u) \int_0^1
dt~t(1-t)
\ln(s_q-t(1-t)\bar{u}-i\epsilon)\right]~.
\end{eqnarray}

     For the first penguin insertion topology, Fig. 2(a), the result is

     \begin{eqnarray}
\overline{\langle Q_1^c\rangle}_{(e)}^{\phi_p} &=& 2\frac{\alpha_s}{4\pi} \frac{C_F}{N}
\langle \pi^- \vert \bar{d}_i  (1+\gamma_5) u_i\vert 0 \rangle
\langle \pi^+ \vert \bar{u}_j (1-\gamma_5) b_j \vert
\bar{B}_d^0 \rangle
\nonumber \\
 &\times&  \left [  \ln \frac{\mu}{m_b}-\frac{1}{2}-3
\int_0^1 du~\phi_p(u) \int_0^1
dt~t(1-t)
\ln(s_c-t(1-t)\bar{u}-i\epsilon)\right]~.
\end{eqnarray}

        Similarly, when $\phi_{\sigma}(u)$ is inserted, by using the method in the
previous subsection, we have

        \begin{eqnarray}
       \langle Q_{4,6}\rangle_{(e)}^{\phi_{\sigma}}&=&\frac{if_{\pi}\mu_{\pi}}{4} g_s^2 \mu^{2\epsilon}
      \frac{C_F}{N} \int_0^1 du ~\frac{\phi_{\sigma}(u)}{6}
      \langle \pi^+\vert \bar{u}_i (1-\gamma_5)
       \gamma_{\alpha}
      \sigma_{\mu\nu} q^{\mu}\gamma_{\rho}
      b_i \vert \bar{B}_d^0 \rangle \nonumber \\
      &\times& \sum \limits_q \left.\left[\frac{\partial}{\partial k_{\nu}}
      \int \frac{d^dl_{q}}{(2\pi)^d}
      \frac{{\rm Tr}[(\slash{\hskip -1.5mm}l_q -\slash {\hskip -2.5mm} k +m_q)
     \gamma^{\alpha} (\slash{\hskip -1.5mm}l_q
      +m_q) \gamma^{\rho}
      (1\mp \gamma_5)]}{[(l_q-k)^2-m_q^2][l_q^2-m_q^2]k^2}\right]
      \right|_{k=p-uq} .
      \end{eqnarray}

        After integration and subtraction,

       \begin{eqnarray}
        \overline{\langle Q_{4,6}\rangle}_{(e)}^{\phi_{\sigma}}&=&
        2\frac{\alpha_s}{4\pi}\frac{C_F}{N}
         \langle \pi^- \vert \bar{d}_i  (1+\gamma_5) u_i\vert 0 \rangle
\langle \pi^+ \vert \bar{u}_j (1-\gamma_5) b_j \vert
\bar{B}_d^0 \rangle \nonumber \\
         &\times&\sum \limits_q \int_0^1 du~ \frac{\phi_{\sigma}}{6\bar{u}}
       \left[ \frac{2}{3}\ln\frac{\mu}{m_b}
         \right. \nonumber \\
        &-&\left. \int_0^1 dt~
        \left(2t(1-t)\ln(s_q-t(1-t)\bar{u}-i\epsilon)+
         \frac{t^2(1-t)^2\bar{u}}{s_q-t(1-t)\bar{u}-i\epsilon}
         \right)\right].
        \end{eqnarray}

         The magnetic penguin insertion is easier; we write the result of
$Q_{8G}$ insertion as follows:

       \begin{eqnarray}
          \langle Q_{8G}\rangle^{\rm twist-3}_{(f)}&=&\frac{\alpha_s}{4 \pi}
         \frac{C_F}{N}f_{\pi}\mu_{\pi} m_b \nonumber \\
         &\times& \int_0^1 du~\left\{\phi_p(u) \frac{1}{k^2}
          \langle \pi^+\vert \bar{u}_i \gamma^{\alpha}\gamma_5\sigma_{\beta\alpha}k^{\beta}
         (1+\gamma_5)b_i\vert \bar{B}_d^0 \rangle  \right.
         \nonumber \\
         &+&i\left. \frac{\phi_{\sigma}(u)}{6}\frac{1}{4k^2}
         \left[\langle \pi^+\vert \bar{u}_i \gamma^{\alpha}\gamma_5
          [\sl q ,\gamma_{\sigma}]~[\gamma^{\sigma},\gamma_{\alpha}](1+\gamma_5)b_i
          \vert  \bar{B}_d^0 \rangle  \right. \right.
         \nonumber \\
         &-&\left.\left.\left. \frac{2}{k^2}
          \langle \pi^+\vert \bar{u}_i \gamma^{\alpha}\gamma_5
          [\sl q ,\sl k]~[\sl k,\gamma_{\alpha}](1+\gamma_5)b_i
          \vert  \bar{B}_d^0 \rangle  \right] \right\}\right |_{k=p-uq}
          \nonumber \\
          &=& 2\frac{\alpha_s}{4\pi}\frac{C_F}{N}(\frac{3}{2}+\int_0^1
          du~\frac{\phi_{\sigma}(u)}{6\bar{u}})
         \langle \pi^- \vert \bar{d}_i  (1+\gamma_5) u_i\vert 0 \rangle
         \langle \pi^+ \vert \bar{u}_j (1-\gamma_5) b_j \vert
          \bar{B}_d^0 \rangle .
         \end{eqnarray}

      \subsubsection{Hard scattering with the spectator}

           The chirally enhanced power corrections in hard spectator scattering
  not only occurs in the case of $(V-A)(V+A)$ vertex insertion, but also in the
case of $(V-A)(V-A)$ insertion. But in the case of $(V-A)(V+A)$ insertion,
after a straightforward calculation, we find
that there will be serious linear divergence at the end points of the
LCDAs if the
twist-3 LCDAs are not symmetric. Because infrared finiteness of the vertex
corrections requires that the twist-3 LCDAs, especially
$\phi_{\sigma}(u)$, must be
symmetric, we shall implicitly assume this symmetric condition for the LCDAs in latter
computation. So, in this symmetric condition,
the hard scattering with the spectator vanishes
when $(V-A)(V+A)$ vertex is inserted. However, even in this strict symmetric
condition, there is still a logarithmic divergence from the end point of
the
recoiling pion in hard spectator scattering
when $(V-A)(V-A)$ vertex is inserted. For example,

           \begin{equation}
\langle Q_2^u \rangle_{(g)+(h)} =  i \pi \alpha_s f_B f_{\pi}^2 \frac{C_F}{N^2}
\int_0^1 d\xi ~\frac{\phi_B(\xi)}{\xi}~\int_0^1 du ~\frac{\phi(u)}{u}~\int_0^1 dv ~
\left[\frac{\phi(v)}{\bar{v}}+\frac{2\mu_{\pi}}{m_B}\frac{\phi_{\sigma}(v)}{6\bar{v}^2}
\right].
\end{equation}

    This means that QCD factorization is broken down. But we
    can still give a phenomenological treatment for this hard
    spectator scattering.
    By using the asymptotic form of $\phi_{\sigma}(u)$, we find that there is a divergent
integral over $v$: $\int_0^1 dv~(1/\bar{v})$. In
Refs.\cite{osaka,beneke1}, the authors
prefer to introduce a phenomenological parameterization for this
logarithmic divergence.
They take $\int_0^1 dv~(1/\bar{v})=\ln(m_B/\Lambda_B) + r e^{i\theta}$,
where
$r$ is taken from 3 (realistic) to 6 (conservative), and the phase $\theta$
from $-\pi$ to $\pi$. We shall take similar phenomenological treatment in the
numerical computation below.

   We notice that the above approach of evaluating hard spectator
contribution
is naive. For instance, the scale of hard spectator contribution should
be different from the vertex correction contribution. While it seems
reasonable to take the scale
$\mu \sim {\cal O}(m_b)$ for the vertex correction diagrams to avoid large
logarithm
$\alpha_s \log (\mu/m_b)$, a natural choice  of the scale of hard
spectator contribution may be around ${\cal O}(1~GeV)$ because the
average
momentum squared of the exchanged gluon is about $1~ GeV^2$.
Another disturbing feature of hard spectator contribution is that, as
pointed out in
Refs. \cite{osaka,beneke1}, when including the contribution of
$\phi_{\sigma}$,
there would appear a divergent integral $\int_0^1 dv (1/\bar{v})$
even if
the symmetric distribution amplitude is applied. This divergent integral
implies
that the dominant contribution comes from the end-point region, or, in
other words, it is dominated by soft gluon exchange. However, the
transverse momentum may not be omitted in the end-point region
\cite{huang}; if so, the
corresponding divergent integral would then be changed to
\begin{equation}
\int dv ~\frac{1}{\bar{v}} \to
\int dv~ d^2 k_T \frac{\Psi(v,k_T)}{\bar{v} \xi m_b^2 +k^2_T}.
\end{equation}
As an illustration, we do not consider the $k_T$ dependence of wave
functions (though it is certainly not a good approximation); then the
above integral is proportional to
\begin{equation}
\int \frac{dv d k^2_T}{\bar{v} \xi m_b^2 +k^2_T} \propto
\int \frac{dx dy}{x+y}.
\end{equation}
The above integration converges now; furthermore it is not dominated
by end-point contribution. This illustrates that the
treatment of hard spectator diagrams may need further
discussion.

There exists ``annihilation topology'' contributions which may belong to
chirally enhanced corrections. In Ref. \cite{beneke1}, the authors have
discussed this topic and find that a divergent integral
$[\int (dx/x)]^2$ will appear. We suspect that this divergence
may disappear, similar to the hard spectator term,  if
the effect of transverse momenta can be included.
It is also possible that ``annihilation topology'' contributions
are really dominated by soft interactions and thus violate factorization.
Due to its complexity, we do not include ``annihilation topology''
contributions in this work.

       \subsection{Final formulas}
       With these effective operators,
     $B \to P_1 P_2$ decay amplitudes
     in QCD factorization can be written as
     \begin{equation}
     A(B\to P_1 P_2)=\frac{G_F}{\sqrt{2}}
     \sum \limits_{p=u,c} \sum \limits_{i=1,10} v_p a^p_i
     \langle P_1 P_2 \vert Q_i \vert $B$ \rangle_F,
     \end{equation}
     where $v_p$ is CKM factor,
     $\langle P_1 P_2 \vert Q_i \vert $B$ \rangle_F$ is the factorized
     matrix element and is the same as the definition of the BSW
    Largrangian\cite{BSW}.
        Then as an illustration, the explicit expressions of $a_i^p$
      ($i=1$ to $10$) for $B \to \pi\pi$ (using symmetric
      LCDAs of the pion) are
      obtained. But it is easy to generalize these formulas to
      the case that the final states are other light pseudoscalars.
      Furthermore, we take only part of QED corrections into account in our final
      formula, in particular the QED penguin insertions.
      Now $a_i^p$ for $B \to \pi \pi$ in NDR $\gamma_5$ scheme is listed as follows
      \footnote{Because of the tedium,
        we do not calculate the radiative corrections in the HV scheme.
       However, generally,
        the results in the NDR scheme and HV scheme can be related
        by a constant matrix $\Delta {\hat r}_s={\hat r}_{s,HV}-{\hat r}_{s,NDR}$
        \cite{buras2}
        which is free from the gauge dependence and infrared structure of the theory.
      Thus, in principle, we can obtain the results in the HV scheme just
       by using $\Delta {\hat r}_s$. In \cite{chy}, the constant matrix has been
       applied to obtain the results in the NDR and HV scheme for the
        coefficients $a_i$ which only contain the current-current vertex corrections.
        But whether we can obtain the expression of $a_6$ or $a_8$ in HV
    scheme in a similar way needs further discussion.}:
      \begin{eqnarray}
      a_1^u&=&C_1+\frac{C_2}{N} + \frac{\alpha_s}{4\pi} \frac{C_F}{N} C_2 F, \\
      a_2^u&=&C_2+\frac{C_1}{N} + \frac{\alpha_s}{4\pi} \frac{C_F}{N} C_1 F,\\
      a_3&=&C_3+\frac{C_4}{N} + \frac{\alpha_s}{4\pi} \frac{C_F}{N} C_4 F, \\
      a_4^p&=&C_4+\frac{C_3}{N} + \frac{\alpha_s}{4\pi} \frac{C_F}{N} C_3 F
      \nonumber \\
      & &- \frac{\alpha_s}{4\pi} \frac{C_F}{N} \left \{
      C_1 (\frac{4}{3}\log\frac{\mu}{m_b}+G(s_p)-\frac{2}{3})+
      (C_3-\frac{C_9}{2})(\frac{8}{3}\log\frac{\mu}{m_b}+G(0)+G(1)-\frac{4}{3})
      \right. \nonumber \\
      & & +\sum_{q=u,d,s,c,b}
      (C_4+C_6+\frac{3}{2}{\rm e_q}C_8+\frac{3}{2}{\rm e_q}C_{10}) \left.
      (\frac{4}{3}\log\frac{\mu}{m_b}+G(s_q))+G_8 C_{8G} \right \}, \\
      a_5&=&C_5+\frac{C_6}{N}+\frac{\alpha_s}{4\pi}\frac{C_F}{N} C_6(-F-12),\\
      a_6^p&=&C_6+\frac{C_5}{N} -\frac{\alpha_s}{4\pi} \frac{C_F}{N}6C_5
      \nonumber \\
      & &-\frac{\alpha_s}{4\pi} \frac{C_F}{N} \left \{
      C_1 ((1+\frac{2}{3}A_{\sigma})\log\frac{\mu}{m_b}-\frac{1}{2}-
      \frac{1}{3}A_{\sigma}+G^{\prime}(s_p)+G^{\sigma}(s_p))
      \right. \nonumber \\
      & &+
      \sum_{q=d,b}(C_3-\frac{C_9}{2})
      ((1+\frac{2}{3}A_{\sigma})\log\frac{\mu}{m_b}-\frac{1}{2}-
      \frac{1}{3}A_{\sigma}+G^{\prime}(s_q)+G^{\sigma}(s_q)) \nonumber \\
      & &+\sum_{q=u,d,s,c,b}
      (C_4+C_6+\frac{3}{2}{\rm e_q}C_8+\frac{3}{2}{\rm e_q}C_{10})
      \left( (1+\frac{2}{3}A_{\sigma})\log\frac{\mu}{m_b}
      +G^{\prime}(s_q)+G^{\sigma}(s_q) \right) \nonumber \\
      & & \left. +(\frac{3}{2}+A_{\sigma})C_{8G} \right \}, \\
      a_7&=&C_7+\frac{C_8}{N}+\frac{\alpha_s}{4\pi}\frac{C_F}{N} C_8(-F-12), \\
      a_8^p&=&C_8+\frac{C_7}{N} -\frac{\alpha_s}{4\pi} \frac{C_F}{N}6C_7
      \nonumber \\
      & &-\frac{\alpha_{em}}{9\pi} \left \{
      (C_2+\frac{C_1}{N})
      ((1+\frac{2}{3}A_{\sigma})\log\frac{\mu}{m_b}-\frac{1}{2}-
      \frac{1}{3}A_{\sigma}+G^{\prime}(s_p)+G^{\sigma}(s_p))
      \right. \nonumber \\
      & &+
      (C_4+\frac{C_3}{N}) \sum_{q=d,b} \frac{3}{2}{\rm e_q}
      ((1+\frac{2}{3}A_{\sigma})\log\frac{\mu}{m_b}-\frac{1}{2}-
      \frac{1}{3}A_{\sigma}+G^{\prime}(s_q)+G^{\sigma}(s_q)) \nonumber \\
      & &+(C_3+\frac{C_4}{N}+C_5+\frac{C_6}{N}) \sum_{q=u,d,s,c,b}
      \frac{3}{2}{\rm e_q}
      \left( (1+\frac{2}{3}A_{\sigma})\log\frac{\mu}{m_b}
      +G^{\prime}(s_q)+G^{\sigma}(s_q) \right) \nonumber \\
      & & \left.
      +(\frac{3}{4}+\frac{1}{2}A_{\sigma})C_{7\gamma} \right \}, \\
      a_9&=&C_9+\frac{C_{10}}{N}+\frac{\alpha_s}{4\pi} \frac{C_F}{N} C_{10} F, \\
      a_{10}^{p}&=&C_{10}+\frac{C_9}{N}+
      \frac{\alpha_s}{4\pi} \frac{C_F}{N}C_{9} F
      - \frac{\alpha_{em}}{9\pi} \left \{
      (C_2+\frac{C_1}{N}) (\frac{4}{3}\log\frac{\mu}{m_b}+G(s_p)-\frac{2}{3})
      \right. \nonumber \\
      & &+(C_4+\frac{C_3}{N})\sum_{q=d,b} \frac{3}{2}{\rm e_q}
      (\frac{4}{3}\log\frac{\mu}{m_b}+G(s_q)-\frac{2}{3}) \nonumber \\
      & & +(C_3+\frac{C_4}{N}+C_5+\frac{C_6}{N}) \sum_{q=u,d,s,c,b} \left.
      \frac{3}{2}{\rm e_q}
      (\frac{4}{3}\log\frac{\mu}{m_b}+G(s_q))+\frac{1}{2}G_8 C_{7\gamma}
      \right \}.
      \end{eqnarray}
      Here $N=3$ is the number of color,
      $C_F=(N^2-1)/2N$ is the factor of color,
      $s_q=m_q^2/m_b^2$ and we define the other symbols
      in the above expressions as
      \begin{eqnarray}
      &&F=-12 \ln \frac{\mu}{m_b} -18+f^{I}+f^{II}, \\
      &&f^{I}=\int \limits_{0}^{1}~ dx~g(x)\phi(x),
      ~G_8=\int \limits_{0}^{1}~ dx~G_8(x) \phi(x), \\
      &&G(s)=\int \limits_{0}^{1}~ dx~G(s,x) \phi(x), \\
      &&G^{\prime}(s)=\int \limits_{0}^{1}~ dx~G^{\prime}(s,x) \phi_p(x),\\
      &&G^{\sigma}(s)=\int \limits_{0}^{1}~ dx~G^{\sigma}(s,x)
      \frac{\phi_{\sigma}(x)}{6(1-x)},~~~~~
      A_{\sigma}=\int \limits_{0}^{1}~ dx~ \frac{\phi_{\sigma}(x)}{6(1-x)},
      \end{eqnarray}
      where $\phi(x)$ [$\phi_p(x)$, $\phi_{\sigma}(x)$] is leading twist
(twist-3)
      LCDA of the ejected pion, and the hard-scattering functions are
      \begin{eqnarray}
      &&g(x)=3 \frac{1-2x}{1-x} \ln x - 3 i \pi, ~~G_8(x)=\frac{2}{1-x}, \\
      &&G(s,x)=-4 \int \limits_{0}^{1}~ du~u(1-u) \ln (s-u(1-u)(1-x)-i
      \epsilon), \\
      &&G^{\prime}(s,x)=-3 \int \limits_{0}^{1}~ du~u(1-u) \ln (s-u(1-u)(1-x)-i
      \epsilon), \\
      &&G^{\sigma}(s,x)=-2 \int \limits_{0}^{1}~ du~u(1-u) \ln (s-u(1-u)(1-x)-i
      \epsilon) \nonumber \\
      &&~~~~~~~~~~~~~~~
      + \int \limits_{0}^{1}~ du~ \frac{u^2(1-u)^2(1-x)}{s-u(1-u)(1-x)-i \epsilon}.
      \end{eqnarray}
      The contributions from the hard spectator scattering [Figs.
1(g),(1)(h)]
      are reduced to the factor $f^{II}$:
       \begin{equation}
      f^{II}=\frac{4 \pi^2}{N}
      \frac {f_{\pi}f_B}{F^{B\to \pi}_{+}(0) m_B^2}
      \int \limits_{0}^{1}~ d\xi~ \frac{\Phi_B(\xi)}{\xi}
      \int \limits_{0}^{1}~ dx~ \frac{\phi(x)}{x} \int \limits_{0}^{1}~
      dy~ \left [ \frac{\phi(y)}{1-y}+\frac{2 \mu_{\pi}}{M_B}
      \frac{\phi_{\sigma}(y)}{6(1-y)^2} \right ].
      \end{equation}

      There contains a divergent integral in $f^{II}$.  Here we simply
      assume that $\int (dy/y) \sim ln (m_b/\Lambda_{QCD})$
(similar to
      what has been done in Refs. \cite{osaka,beneke1}, though our
assumption here is
      certainly an oversimplification). We thus illustrate numerically the
      scale dependence of $a_i^p(\pi\pi)$ in Table 1. Here we use the
      asymptotic form of the LCDAs of the light pseudoscalar meson which are
      listed in Appendix A,
      and the other input parameters are taken as follows\cite{ali}:
      $F^{B\pi}(0)=0.33$, $f_B=0.2~GeV$, $f_{\pi}=133~MeV$,
      the pole masses $m_b=4.8~GeV$, $m_c=1.4~GeV$, the $\overline {MS}$ masses
      ${\overline m_t}({\overline m_t})=170~GeV$,
      ${\overline m_b}({\overline m_b})=4.4~GeV$,
      $\overline {m}_u(2~GeV)=4.2~MeV$, $\overline{m}_d(2~GeV)=7.6~MeV$ and
      $\Lambda_{QCD}^{(5)}=225~MeV$.

           \section{Discussions and General Remarks}
            \subsection{Color transparency and factorization}

        Color transparency gives a clear physics picture of QCD
        factorization. In the argument of the color transparency,
        a fast-moving small color singlet formed by a pair of $q\bar{q}$
         decouples from the surrounding soft gluons. Of course,
        as mentioned in the previous section,
        only the decoupling with the soft gluons is not enough for
        a factorization formula, the decoupling from the collinear
        divergence is also necessary. We really find that both of the requirements
        can be satisfied in the one-loop calculations. So
        the QCD factorization is guaranteed. Therefore, the
        calculations in the above sections seem to be a one-loop
        technical manifestation (or demonstration) of the color
        transparency. On the other hand, at the leading power of
        $\Lambda_{QCD}/m_b$, the soft or collinear gluons only ``see'' the
        direction of the light meson, but are ``blind'' to the spins of
        the quark constituents.
        So the soft or collinear gluon cannot distinguish whether the ejected meson
        from $b$ quark decay is a light pseudoscalar or a light longitudinally polarized
        vector meson. As a consequence, the cancellation of the infrared divergence is
        universal for $B$ decays to two light mesons, no matter whether the
        meson is a
        pseudoscalar meson or a vector meson. Therefore, the QCD factorization formula
        for $B \to PP$ at the leading
        power of $\Lambda_{QCD}/m_b$ is easy to be generalized to $B \to
        PV$ and $VV$.

        Similarly, the  color transparency
        argument can not only be applied to the strong
        interactions, but also be generalized to the electromagnetic
        interactions.
       When the ejected meson is electric neutral,
         the soft photons also decouple from the fast moving
         small electric dipole. So QED vertex corrections
             are also of infrared finiteness.
             But for the case that the
      ejected meson is charged,
      QED corrections are infrared divergent, and the
      infrared divergence must be cancelled by the soft photon emission
      mechanism, which is common in the calculation of the radiative corrections
      for $\mu^- \to e^- \nu_{\mu} \bar{\nu}_e$. About this, it is easy to be
      covered in the calculation in the previous section, just replacing
the QCD vertex
      by a QED vertex. This can be called a one-loop demonstration
      for ``charge'' transparency.

     It should be noted that the above arguments must be based on
     the condition that the ejected meson is in a very
     compact configuration, then it, as
     a small color dipole, is disentangled with the soft gluons.
      Otherwise, if its size is too large, it is difficult to decouple from the
      soft gluons. For example, the spectator quark in D meson
      is very soft, and runs around c quark like a soft quark cloud,
    which has a large overlap with the $B$ meson spectator system\cite{nucl}.
    As a consequence, the process in which D meson is ejected from $b$ decay is
    dominated by the soft gluon exchange.

            \subsection{The scale dependence}
         From the expressions of the QCD coefficients $a_i^p$
 obtained in previous sections, it is found that the
 renormalization scale dependence of the hadronic matrix elements of the
 effective operators is recovered. Apparently, we expect this
 recovered dependence can cancel the scale dependence of the Wilson
 coefficients $C_i$.

           With the renormalization group equations for the Wilson
      coefficients $C_i(\mu)$ at leading order logarithm approximation\cite{buras},

      \begin{equation}
      \mu \frac{d }{d \mu}{\bf C(\mu)}=\frac{\alpha_s}{4\pi}
\hat{\gamma}^{(0)T}
      {\bf C(\mu)}
      \end{equation}
     we do find
                  \begin{equation}
                   \mu \frac{d }{d\mu}a_i^p =0
                    ~~~~~~~~~~~~({\rm for~
                     i=1-5 ~and~ 7,~ 9, ~10)}
                   \end{equation}
       when we neglect the contributions from higher order of $\alpha_s$.
       But for $a_{6}$ or $a_{8}$, some scale
      dependence at the order of $\alpha_s$ still remains. Note that
      other QCD coefficients ($a_{i=1,2,3,4,5,7,9,10}$) are multiplied by the
      product of the matrix elements
      of the conserved currents which are independent of the
      renormalization scale;
      whereas the coefficient $a_6$ or $a_8$ is multiplied by
       a product of the
       two matrix elements of scalar and pseudoscalar current
      \[
      -2\langle P_1 \vert \bar{d} (1+\gamma_5) q \vert 0 \rangle
       \langle P_2 \vert \bar{q} (1-\gamma_5) $b$ \vert $B$ \rangle.
       \]
     which is still of scale dependence. This scale dependence is
     generally represented by the factor
      \[
      r_{\chi}(\mu)=\frac{2m_{P_1}^2}{\overline{m}_b(\mu)
      (\overline{m}_1(\mu)+\overline{m}_2(\mu))}
      \]
      after we apply the equations of motion to transform the $(S+P)(S-P)$ matrix
     elements into the type of $(V-A)(V-A)$.
     Here $m_1$ and $m_2$ are the current masses of the valence
      quarks in meson $P_1$.
      With the renormalization group equations for the running
      mass of the current quark
      \begin{equation}
      \mu \frac{d }{d\mu}\overline{m}(\mu)
      =-6\frac{\alpha_s}{4 \pi} C_F \overline{m}(\mu),
      \end{equation}
      we have
      \begin{equation}
      \mu \frac{d}{d\mu}r_{\chi}(\mu)=12\frac{\alpha_s}{4 \pi} C_F
      r_{\chi}(\mu)~.
      \end{equation}
      Consequently, we find
      \begin{equation}
     \mu  \frac{d}{d\mu} \left(a_{6,8}^p r_{\chi}\right)=0~
      \end{equation}
        with the asymptotic form of $\phi_{\sigma}(u)=6u(1-u)$.
        Then, as we expect, the decay amplitude for $B$ decays to two light
        pseudoscalars predicted by QCD factorization is really
        independent of the renormalization scale within the constraint
         $A_{\sigma}=1/2$. This also can be
       obviously seen from the numerical results of $a_i^p(\pi \pi)$
listed in Table 1.
       In particular, if we think that the results of QCD factorization are reliable
       and really independent of the renormalization scale, maybe $A_{\sigma}=1/2$
       is a strict constraint for the form of $\phi_{\sigma}(u)$.

       It should be noted
         that the imaginary part of QCD coefficients $a_i$ only arises at
         the order of $\alpha_s$, and depends on the renormalization
         scale. This dependence would bring some uncertainties in determining
         the $CP$ asymmetries in $B$ decays. Maybe this scale dependence of the
         imaginary part could be cancelled by the results on higher order of
        $\alpha_s$.

       \subsection{Comparison to the generalized
       factorization and PQCD method}
         Comparing QCD factorization approach with the generalized
      factorization\cite{ali,chy3} and PQCD method\cite{lihn2}, some
      interesting comments are in order.

      (i) At the zeroth order
      of $\alpha_s$, both of QCD factorization (BBNS approach) and generalized
      factorization can reproduce the results of ``naive factorization'';
at
      the higher order of $\alpha_s$, the renormalization scheme and scale
      dependence for the hadronic matrix elements can be recovered from the
      hard-scattering kernels $T^{I}_{i}$ in BBNS approach and $\hat{\bf m}_s$
      in generalized factorization.
      However, in generalized factorization, $\hat{\bf m}_s$ is from the
      one-loop calculations of quark-level matrix elements. According to Buras
      {\it et al.}, quark-level matrix elements are accompanied with infrared
      divergences. To avoid these divergences, one may assume that external
      quarks are off-shell. Unfortunately, it will introduce gauge
      dependence which is also unphysical. But in the BBNS approach, it is
different
      because the external states are all physical and can be approximated as
      on-shell quarks in the leading order of $\Lambda_{QCD}/m_b$.
      As a consequence, the unphysical gauge dependence
      does not appear. In the PQCD method, Li {\it et al.} claim that
their
      method is based on a six-quark system in which the external quarks
are
      on-shell, so the gauge invariance of PQCD predictions is guaranteed.
        The scale dependence in PQCD prediction is removed by evolving the Wilson
        coefficients down to the proper hard scale.
      So no explicit scale dependence is left.

      (ii) In generalized factorization and BBNS approach, the hadronic transition
      form factors are not calculable, and they are dominated by
      soft gluon exchange and determined only by experiments or some non-perturbative
      approaches such as sum rule, lattice QCD, etc. In particular,
the above
      assumption can be also justified in the BBNS approach by naive
power counting
      \cite{beneke,nucl}. However, in the PQCD method, this naive power
counting rule
      may be invalid when the transverse momenta of the quark constituents and
      Sudakov suppression are taken into account.
      So, Li {\it et al.} thought that
      the hadronic transition form factors can be calculated in the PQCD
method because
      $b$ quark
      is heavy enough and the soft gluon exchange is suppressed by Sudakov
      form factors. This is the essential difference
      between the BBNS approach and the PQCD method.

      (iii) The generalized factorization considers ``nonfactorizable''
contributions as
      intractable. Therefore, one may introduce one or more effective color numbers
      $N_c^{eff}$ to phenomenologically represent ``nonfactorizable''
contributions
      \cite{chy1,ali,chy3}.
      Furthermore, $N_c^{eff}$ is assumed to be universal to maintain predictive power.
      However, ``nonfactorizable'' contribution is really
process-dependent.
      In the BBNS approach and the PQCD method, such ``nonfactorizable''
contributions
      are indeed calculable in perturbative theory.
      In consequence, $N_c^{eff}$ need not be introduced.

      (iv) As mentioned in the above sections, the strong phases predicted by generalized
      factorization are only from the BSS mechanism which is represented
by the penguin
      insertion. However, the virtuality of the gluon or photon $k^2$ in the penguin insertion
      is ambiguous in generalized factorization, and usually it is approximated around
      $m_B^2/2$. This brings significant uncertainties for predicting the $CP$ asymmetries for B
      decays. A particular interesting result of the BBNS approach is that
strong
      phases are not only from the BSS mechanism but also from the hard
scattering,
      and there are no
      uncertainties in determining $k^2$ of penguin insertion.
      However, compared with the real part of the decay amplitude, the imaginary part
      is ${\cal O} (\alpha_s)$ or
      power $\Lambda_{QCD}/m_b$ suppressed and cannot lead to large $CP$ asymmetries,
      since they come solely from
      hard scattering processes which are only calculable in the heavy quark limit.
      In the PQCD method, there is no such $\alpha_s$ suppression in the
imaginary part of
      the decay amplitude.
      Thus $CP$ asymmetries predicted by the PQCD method are usually greater
than the
      prediction of BBNS approach and generalized factorization.
      So maybe these differences of prediction for $CP$ asymmetries can be
an
      experimental test for the BBNS and PQCD approaches.

      (v) Hard spectator contributions [Fig. 1(g)and 1(h)], which are
leading power
      effects in QCD factorization, miss out in ``naive factorization'' and
      ``generalized factorization''. They are, however, ${\cal O}(\alpha_s)$ suppressed
       compared with the leading factorized contributions (the hadronic transition
       form factors). But, in the PQCD method,
       they are of the same order as the form factors.

      (vi) In the PQCD method, penguin contributions receive a dynamical
enhancement
       called ``Fat Penguin''\cite{lihn}. But in generalized and QCD factorization, they
       are missed. This enhancement in the PQCD method arises from the
strong scale dependence
       of the penguin Wilson coefficients $C_{4,6}$, etc.

     (vii) Final state interactions (FSI) do not appear in the three
     methods. In QCD factorization, Beneke {\it et al.} point out that the cancellation
      of the infrared divergences implies that the long distance FSI is power suppressed
      due to the quark-hadron duality. However, this point of view is controversial
      \cite{fsi}, but can be examined by the experimental measurements of $B \to
       KK$\cite{lihn3}.

          \subsection{Limitation of QCD factorization}

             QCD factorization formula only holds in the heavy quark limit
         $m_b \to \infty$. In the real world $m_b$ is only about $4.8$
GeV, the
         validity of the power suppression may be questionable. In particular, for
         several cases, the power suppressed corrections can be numerically
         large\cite{nucl}, because the perturbative expansion is in order
of $\alpha_s$
         which is not small at the realistic scale
         ${\cal O}(m_b)$ compared to $\Lambda_{QCD}/m_b$.

         (i) The hard ``nonfactorizable'' contributions computed by QCD
factorization
         are generally small compared to the leading ``factorizable''
contribution.
        But when the leading ``factorizable'' contributions are color
suppressed,
         the ``nonfactorizable'' contribution may be larger than the
leading results.
         At the same time, the potentially soft contribution, which is formally power
         suppressed, may be important.
         For example, in $\bar{B}_d^0 \to \pi^0 \pi^0$, any perturbative and
         soft power suppressed contributions can have a significant
effect on
         predicting the branching ratio and $CP$ asymmetry. Furthermore, this problem
         also arises when the entire leading power contribution is suppressed by small
         Wilson coefficients, for example, in $B \to KK$;
          or when the leading power
         contribution is suppressed by the small CKM elements.

          (ii) An important power suppressed contribution is from the higher twist
         light-cone wave functions of the light mesons. The chirally enhanced power
         correction from the two-particle twist-3 wave functions is the most important,
         and has been partly involved in this work except for the
         annihilation topologies.
         Other contribution from multiparticle non-valence
        fock state has been proved to be also power suppressed\cite{nucl}. However, there
        is no systematic way to evaluate it. The author of Ref.\cite{Kho} proposed a
        way to evaluate the soft gluon exchange contribution from higher twist
         $q\bar{q}g$ wave functions within the frame of the light-cone sum
rule(LCSR).
         But the accuracy of LCSR is limited due to the quark-hadron duality
        approximation. On the other hand, power correction
         from transverse momenta needs a subtle treatment in the future.
          In Ref.\cite{nucl}, the authors
         point out that the contribution from the transverse momenta might be
         considered when we evaluate the hadronic matrix elements to two-loop
         order. In this case, it is possible that Sudakov suppression might be
         taken into account as well.

           In summary, up to now, we do not have a systematic way to evaluate
         many kinds of power suppressed corrections for exclusive processes.
         How to evaluate
         such corrections in a consistent way within the frame of QCD factorization is
         a potentially interesting work.

                 \section{summary}

               In this work, we give a detailed discussion for QCD factorization
         involving the complete chirally enhanced power corrections
       in the heavy quark limit for $B$ decays to two light pseudoscalar mesons, and present some
       elaborate calculations
       of radiative corrections at the order of $\alpha_s$. We point
       out that the infrared finiteness of the vertex corrections
       in the chirally enhanced power corrections requires
       twist-3 light-cone distribution amplitudes (LCDAs) of the light
       pseudoscalar symmetric. However, even in the symmetric condition,
       there is also infrared divergence from the end point of the
       LCDAs in the hard spectator scattering and annihilation topology.
       So the transverse momenta and Sudakov
        suppression should be taken into account. We also point out that
       the decay amplitude of $B\to PP$ predicted by QCD factorization is really
        independent of the renormalization scale, at least at the order
        of $\alpha_s$.  At last, we briefly compare the QCD
        factorization to the generalized factorization and PQCD
        method which are generally used in studying $B$ exclusive hadronic decays.

        \begin{center}
       {\bf Acknowledgements}
        \end{center}

          We thank Professor Hai-Yang Cheng and Kwei-Chou Yang for
pointing out
          errors in the coefficients of $C_{8G}$ and $C_{7\gamma}$ in
Eqs.(50) and (52).
          And we also thank Professor Maozhi Yang for
          helpful discussions about QCD factorization and the PQCD
          method.
          This work is supported in part by National Science
          Foundation of China and State Commission of Science and
          Technology of China.

     \begin{center}
     \bf{Appendix A.  Twist-2 and -3 LCDAs of
      Light Pseudoscalar Meson}
       \end{center}

      Two particle twist-2 and twist-3 light-cone distribution amplitudes of
      light pseudoscalar mesons are defined by the following
      nonlocal matrix elements\cite{t3}:
      \begin{eqnarray}
      \langle P(p^{\prime})\vert \bar{q}(y) \gamma_{\mu} \gamma_5
      q(x) \vert 0 \rangle &=& -i f_P p^{\prime}_{\mu} \int_0^1 du
      e^{iup^{\prime}\cdot y+i\bar{u} p^{\prime}\cdot x} \phi(u),
      \\
        \langle P(p^{\prime})\vert \bar{q}(y) \gamma_5
      q(x) \vert 0 \rangle &=& -i f_P\mu_{P} \int_0^1 du
      e^{iup^{\prime}\cdot y+i\bar{u} p^{\prime}\cdot x} \phi_p(u),
      \\
        \langle P(p^{\prime})\vert \bar{q}(y) \sigma_{\mu\nu} \gamma_5
      q(x) \vert 0 \rangle &=& i f_P \mu_P (p^{\prime}_{\mu} z_{\nu}-
      p^{\prime}_{\nu} z_{\mu})\int_0^1 du
      e^{iup^{\prime}\cdot y+i\bar{u} p^{\prime}\cdot x}
      \frac{\phi_{\sigma}(u)}{6},
      \end{eqnarray}
      with $f_P$ being the decay constant of the light pseudoscalar,
      $\mu_P=M_P^2/(m_1+m_2)$ ($m_1$ and $m_2$ are the masses of the
constituent
      quarks in the pseudoscalar), and $z=y-x$. Here $\phi(u)$ is
      the twist-2 light-cone distribution amplitude;
      $\phi_p(u)$ and $\phi_{\sigma}(u)$ are two-particle twist-3
      distribution amplitudes. The above definitions can be
      combined into the below nonlocal matrix element:
      \begin{eqnarray}
      \langle P(p^{\prime}) \vert \bar{q}_{\alpha}(y) q_{\beta}(&x&)
      \vert 0 \rangle = \frac{if_P}{4} \int_0^1
      e^{iup^{\prime}\cdot y+i\bar{u} p^{\prime}\cdot x} \nonumber
      \\
      &&\times  \left \{ \slash{\hskip -2.5mm}p^{\prime} \gamma_5
      \phi(u)-\mu_P \gamma_5 \left(\phi_p(u)-\sigma_{\mu\nu} p^{\prime
      \mu}z^{\nu} \frac{\phi_{\sigma}(u)}{6}\right) \right \}_{\beta\alpha} ~.
      \end{eqnarray}

      Neglecting the three-particle twist-3 light-cone wave function, the
asymptotic forms of the above distribution amplitudes are given as
      \begin{eqnarray}
      \phi(u)&=&6u(1-u), \\
      \phi_p(u)&=&1, \\
      \phi_{\sigma}(u)&=&6u(1-u).
      \end{eqnarray}

      \begin{center}
      {\bf Appendix B. The Evolution of $C_i(\mu)$ }
      \end{center}

      The renormalization group equation for the Wilson coefficients
$C_i(\mu)$ is written as follows\cite{buras}:
      \begin{equation}
      \mu \frac{d}{d\mu} {\bf C}(\mu)=\hat{\gamma}^T {\bf
C}(\mu)
      \end{equation}
      Here $\gamma$ is the anomalous dimension matrix, which can be
calculated by the perturbative theory and expanded in order of the
coupling constants $\alpha_s$ and $\alpha_{em}$:
      \begin{equation}
      \hat{\gamma}=\frac{\alpha_s}{4\pi} \hat{\gamma}_s^{(0)}+
      (\frac{\alpha_s}{4\pi})^2 \hat{\gamma}_s^{(1)}+
      \frac{\alpha_{em}}{4\pi} \hat{\gamma}_e^{(0)}+
       \frac{\alpha_s \alpha_{em}}{(4 \pi)^2} \hat{\gamma}_{se}^{(1)}+
       \cdot \cdot \cdot
      \end{equation}
       The LO anomalous dimension matrix $\gamma_s^{(0)}$ of the above
equations has the explicit form
    \begin{equation}
    \hat{\gamma}_s^{(0)}=\left(
     \begin{array}{cccccccccc}
     \frac{-6}{N} & 6 & \frac{-2}{3N} & \frac{2}{3} & \frac{-2}{3N}
      & \frac{2}{3} & 0 & 0 & 0 & 0 \\
      6 & \frac{-6}{N}& 0 & 0 & 0 & 0 &  0 & 0 & 0 & 0 \\
      0 & 0 & \frac{-22}{3N} & \frac{22}{3} & \frac{-4}{3N} & \frac{4}{3}
      & 0 & 0 & 0 & 0 \\
     0 & 0 & 6-\frac{2f}{3N} & \frac{-6}{N}+\frac{2f}{3} & \frac{-2f}{3N}
     & \frac{2f}{3} & 0 & 0 & 0 & 0 \\
     0 & 0 & 0 & 0 & \frac{6}{N} & -6 & 0 & 0 & 0 & 0 \\
     0 & 0 & \frac{-2f}{3N} & \frac{2f}{3} & \frac{-2f}{3N}
     & \frac{-6(N^2-1)}{N}+\frac{2f}{3} & 0 & 0 & 0 & 0 \\
     0 & 0 & 0 & 0 & 0 & 0 & \frac{6}{N} & -6 & 0 & 0 \\
     0 & 0 & \frac{-2(u-d/2)}{3N} & \frac{2(u-d/2)}{3}
       & \frac{-2(u-d/2)}{3N} & \frac{2(u-d/2)}{3} & 0
       & \frac{-6(N^2-1)}{N}  & 0 & 0 \\
     0 & 0 & \frac{2}{3N} & -\frac{2}{3} & \frac{2}{3N} & -\frac{2}{3} & 0
       & 0 & \frac{-6}{N} & 6 \\
     0 & 0 & \frac{-2(u-d/2)}{3N} & \frac{2(u-d/2)}{3}
       & \frac{-2(u-d/2)}{3N}     &  \frac{2(u-d/2)}{3} & 0 & 0 & 6
       & \frac{-6}{N}
     \end{array}
     \right)
      \end{equation}
     where $N$ is the color number, $f$ is the active flavor
     number, and $u$ and $d$ denote the number of the active up-
     and down-type flavors respectively.

      \begin{center}
      {\bf Appendix C. Some Useful Feynman Parameter Integrals}
      \end{center}
         In calculation of the perturbative diagrams shown in Fig. 1, one
might encounter some Feynman parameter integrals which involve nontrivial
infrared divergence. To deal with the infrared divergence, as
mentioned in preceding sections,
the dimensional regularization (DR) and massive gluon (MG) scheme are applied.
Below, we give the explicit calculation of some useful Feynman parameter
integrals
in the above two regularization schemes.

   First, we deal with the integrals in DR scheme. In the DR scheme (here we take
   $d=4+2a$ and $a>0$), the integrals involving the infrared divergence
are
   written as follows:
  \begin{eqnarray}
    \int_0^1 dt_1 \int_{0}^{1-t_1} dt_2
    \frac{1}{(t_1(t_1+t_2 u))^{1-a}} &=&
     \frac{1}{u}\left[\frac{1}{2a^2}+\frac{\ln u}{a}+\frac{1}{2}\ln^2
    u-{\rm Li}_2(1-\frac{1}{u})\right], \\
    \int_0^1 dt_1 \int_{0}^{1-t_1} dt_2
    \frac{t_2}{(t_1(t_1+t_2 u))^{1-a}}&=&
   \frac{1}{u}\left[\frac{1}{a}-2+\ln u+\frac{\ln u}{1-u}\right], \\
   \int_0^1 dt_1 \int_{0}^{1-t_1} dt_2
   \frac{(1-t_1)(1-t_2)}{(-t_1 t_2 u)^{1-a}}&=&-\frac{1}{u}\left[
   \frac{1}{a^2}+\frac{\ln(-u)-2}{a}+\frac{27-\pi^2}{6}\right.\nonumber \\
   &-& \left. 2\ln(-u)+\frac{1}{2}\ln^2(-u)
   \right] .
    \end{eqnarray}

   Here ${\rm Li}_2(x)$ is the dilogarithm function. It is defined by

   \begin{equation}
   {\rm Li}_2(x)=-\int_0^x \frac{\ln (1-t)}{t} dt~.
   \end{equation}

           The Feynman parameter integrals in the MG scheme
       are listed as follows:

    \begin{eqnarray}
    &&\int_0^1 dt_1 \int_{0}^{1-t_1} dt_2
    \frac{1}{(t_1(t_1+t_2 u)+(1-t_1-t_2) \lambda)}\nonumber \\
     &=&\frac{1}{u}\left[\frac{1}{4}\ln^2 \lambda +\ln(-u)\ln\lambda
      -2 \ln u \ln \lambda+\frac{1}{2}\ln^2 u-
    {\rm Li}_2(1-\frac{1}{u}) + \frac{5}{4}\pi^2\right], \\
    &&\int_0^1 dt_1 \int_{0}^{1-t_1} dt_2
    \frac{t_2}{(t_1(t_1+t_2 \omega)+(1-t_1-t_2) \lambda)}\nonumber \\
    &=&\frac{1}{u}\left[-\ln \lambda-1+\ln u+\frac{\ln u}{1-u}\right], \\
    &&\int_0^1 dt_1 \int_{0}^{1-t_1} dt_2
    \frac{(1-t_1)(1-t_2)}{-t_1 t_2 u+(1-t_1-t_2)\lambda}\nonumber \\
    &=&-\frac{1}{u}\left[
   \frac{1}{2}\ln^2\lambda-(\ln(-u)-2)\ln \lambda
   - 2\ln(-u)+\frac{1}{2}\ln^2(-u)+\frac{5}{2}+\frac{\pi^2}{3}
   \right] .
    \end{eqnarray}

    When we calculate the above integrals in the MG scheme, the following
equations
about dilogarithm function may be useful:

    \begin{eqnarray}
    {\rm Li}_2(-x)+{\rm Li}_2(-\frac{1}{x})&=&
    -\frac{\pi^2}{6}-\frac{1}{2}\ln ^2 x ~~~(x>0) \\
     {\rm Li}_2(x)+{\rm Li}_2(\frac{1}{x})&=&
     \frac{\pi^2}{3}-\frac{1}{2}\ln ^2 x -i\pi \ln x~~~(x>1) \\
    {\rm Li}_2(ix)+{\rm Li}_2(-\frac{i}{x})&=&
     -\frac{\pi^2}{24}-\frac{1}{2}\ln ^2 x
    +\frac{i}{2}\pi \ln x~~~(x>0) \\
    {\rm Li}_2(-ix)+{\rm Li}_2(\frac{i}{x})&=&
    -\frac{\pi^2}{24}-\frac{1}{2}\ln ^2
     x-\frac{i}{2}\pi \ln x ~~~(x>0) \\
      {\rm Li}_2(x)+{\rm Li}_2(1-x)&=&\frac{\pi^2}{6}-\ln x \ln (1-x)
    \end{eqnarray}

           \begin{table}
     \vspace*{1cm}
     \begin{tabular}{ccccc}
     QCD &  \multicolumn{2}{c} {$\mu=5.0~GeV$} &
     \multicolumn{2}{c} {$\mu=2.5~GeV$} \\
     Coefficients & NLO & LO  & NLO & LO \\ \hline
     $a_1^u$ & $1.024+0.012 i$   & $1.017$  &
               $1.034+0.024 i$   & $1.037$  \\
     $a_2^u$ & $0.144-0.076 i$ & $0.188$  &
               $0.123-0.100 i$ & $0.109$  \\ \hline
     $a_3$   & $0.003+0.002 i$ & $0.002$  &
               $0.004+0.004  i$ & $0.004$   \\
     $a_4^u$ & $-0.027-0.014 i$ & $-0.029$ &
               $-0.029-0.017 i$ & $-0.040$ \\
     $a_4^c$ & $-0.033-0.007 i$ & $-0.029$ &
               $-0.036-0.007 i$ & $-0.040$  \\
     $a_5$   & $-0.003-0.003 i$ & $-0.005$ &
               $-0.002-0.005 i$ & $-0.010$ \\
     $r_{\chi} a_6^u$ & $-0.036-0.012 i$ & $-0.033$ &
               $-0.037-0.011 i$ & $-0.040$ \\
     $r_{\chi} a_6^c$ & $-0.039-0.005 i$ & $-0.033$   &
               $-0.040-0.004 i$ & $-0.040$ \\ \hline
     $a_7 \times 10^{5}$   & $11.9+2.8 i$     & $13.8$   &
               $0.0+5.4 i$     & $7.6$  \\
     $r_{\chi} a_8^u \times 10^{5}$ & $36.8-10.9 i$     & $36.8$   &
               $45.0-5.2 i$     & $39.8$ \\
     $r_{\chi} a_8^c \times 10^{5}$ & $35.0-6.2 i$     & $36.8$   &
               $44.2+3.1 i$     & $39.8$ \\
     $a_9 \times 10^{5}$   & $-936.1-13.4 i$  & $-928.4$ &
               $-953.9-24.5 i$  & $-957.3$ \\
     $a_{10}^u \times 10^{5}$ & $-81.8+58.8 i$   & $-141.4$ &
                $-58.3+86.1 i$   & $-74.0$ \\
     $a_{10}^c \times 10^{5} $ & $-85.2+63.5 i$   & $-141.4$ &
                $-60.3+88.8 i$   & $-74.0$ \\
     \end{tabular}

     \vspace{0.5cm}
     \caption{ The QCD coefficients $a_i^p(\pi \pi)$ at NLO and LO for the
     renormalization scales at $\mu=5~ GeV$ and $\mu=2.5~GeV$, where
     $r_{\chi}=2 m_{\pi}^2/m_b (m_u+m_d)$. }

     \end{table}

   \begin{figure}[tb]
     \vspace*{1cm}
     \centerline{\epsfig{figure=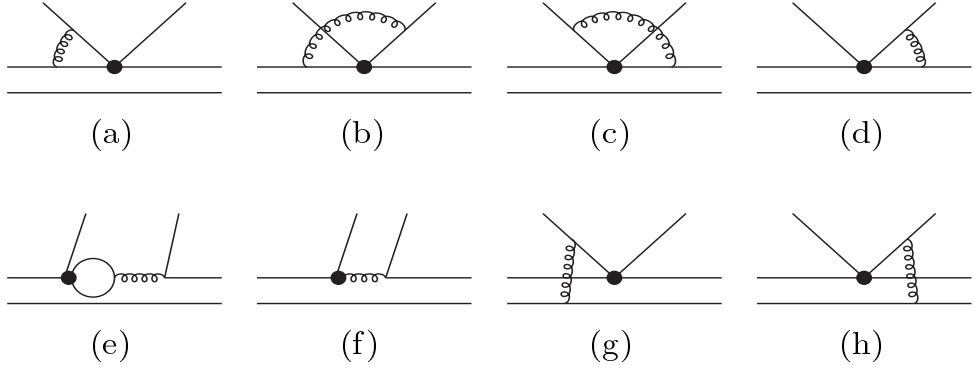,height=6cm,width=15cm,angle=0}}
     \vspace*{1.cm}
     \caption{Order of $\alpha_s$ corrections to hard-scattering kernels $T^I$
     and $T^{II}$. The upward quark lines represent the ejected quark pairs
     from $b$ quark weak decays.}
     \end{figure}

       \begin{figure}[tb]
       \vspace*{1cm}
       \centerline{\epsfig{figure=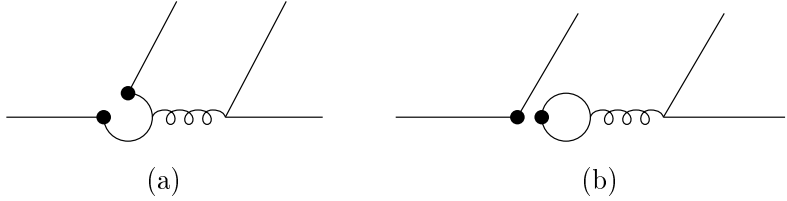,height=3cm,width=11cm,angle=0}}
       \vspace*{1.cm}
       \caption{ Two kinds of topology for penguin contractions. }
       \end{figure}

       \begin{figure}[tb]
       \vspace*{1cm}
       \centerline{\epsfig{figure=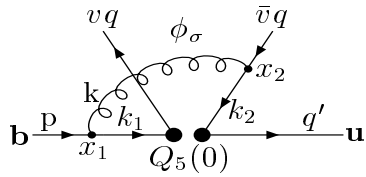,height=3cm,width=5cm,angle=0}}
       \vspace*{.5cm}
         \caption{An example of the vertex corrections for the operator $Q_5(0)$ in
coordinate space in the case of $\phi_{\sigma}$ insertion. }
       \end{figure}

            \end{document}